\documentclass{IEEEtran}
\usepackage{cite}
\usepackage{amsmath,amssymb,amsfonts}
\usepackage{graphicx}
\usepackage{textcomp,nicefrac}
\usepackage[version=4]{mhchem} 
\usepackage{booktabs}
\usepackage{multirow}
\def\BibTeX{{\rm B\kern-.05em{\sc i\kern-.025em b}\kern-.08em T\kern-.1667em\lower.7ex\hbox{E}\kern-.125emX}}
\usepackage{subfigure,dcolumn}
\usepackage{epstopdf}
\usepackage{mhchem}
\usepackage{upgreek}
\usepackage{hyperref}
\usepackage{url}

\begin{document}
\title{Characterization of a Gaseous Time Projection Chamber With an Internal \ce{^{37}Ar} Source}

\author{Wenming Zhang, Yuanchun Liu, Ke Han, Shaobo Wang, Xiaopeng Zhou, and Xunan Guo 
\thanks{Manuscript received xx xx xx; accepted xx xx xx. Date of publication xx xx xx; date of current version xx xx xx.}
\thanks{This work is supported by grant No.U1965201 and No.11905127 from the National Natural Science Foundation of China and grant 2016YFA0400302 from the Ministry of Science and Technology of China. We thank the support from the Key Laboratory for Particle Physics, Astrophysics and Cosmology, Ministry of Education. We thank the support from the Double First Class Plan of Shanghai Jiao Tong University. (Wenming Zhang and Yuanchun Liu contributed equally to this work.) (Corresponding authors: Ke Han and Shaobo Wang.)}
\thanks{Wenming Zhang, Yuanchun Liu, and Ke Han are with the INPAC and School of Physics and Astronomy, Shanghai Jiao Tong University, MOE Key Lab for Particle Physics, Astrophysics and Cosmology, Shanghai Key Laboratory for Particle Physics and Cosmology, Shanghai 200240, China (e-mail: ke.han@sjtu.edu.cn).}
\thanks{Shaobo Wang is with the INPAC and School of Physics and Astronomy, Shanghai Jiao Tong University, MOE Key Lab for Particle Physics, Astrophysics and Cosmology, Shanghai Key Laboratory for Particle Physics and Cosmology, Shanghai 200240, China. He is also with the SPEIT (SJTU-Paris Elite Institute of Technology), Shanghai Jiao Tong University, Shanghai 200240, China (e-mail: shaobo.wang@sjtu.edu.cn).}
\thanks{Xiaopeng Zhou and Xunan Guo are with the School of Physics, Beihang University, Beijing 102206, China.}}

\maketitle

\begin{abstract}
We report on a novel calibration method of gaseous detectors using an internal \ce{^{37}Ar} source.
The \ce{^{37}Ar} is a fast-decaying and low-energy calibration source that provides a mono-energetic peak of 2.82 keV. 
A gaseous \ce{^{37}Ar} source is injected and uniformly distributed in a Micromegas-based gaseous time projection chamber (TPC).
Key performance parameters of the detector, such as electron transmission, gain, energy resolution, gain uniformity, and drift field evolution, are effectively and quickly calibrated.
The gain uniformity, related to the homogeneity of the avalanche gap of Micromegas, is calibrated quickly thanks to the event-by-event position reconstruction and quasi-point energy deposition of \ce{^{37}Ar}.
The energy resolution is improved with the obtained gain uniformity map.
The most noticeable improvement in energy resolution, from 44.9\% to 35.4\%, is observed at a working pressure of 7 bar.
The internal calibration source is also used to characterize the dependence of the detector's electric field distortion on the drift field.
\end{abstract}


\section{Introduction}\label{sec.I}

\IEEEPARstart{T}{ime} Projection Chambers (TPCs), first introduced in the 1970s~\cite{bib:1}, have become widely utilized in nuclear and particle physics experiments.
Gaseous TPCs can record both energy depositions and three-dimensional trajectories of charged particles traveling within, providing information to discriminate the signal from the background~\cite{bib:2, bib:3}.
They have applications in collider experiments (e.g., ALICE~\cite{bib:4}, STAR~\cite{bib:5}), Neutrinoless Double Beta Decay (NLDBD) experiments (e.g., NEXT~\cite{bib:6}, NvDEx~\cite{bib:7}), dark matter detection experiments (e.g., TREX-DM~\cite{bib:8}, NEWS-G~\cite{bib:9}, MIMAC~\cite{bib:10}), and low-radioactive material screening~\cite{bib:11},~\cite{bib:12},~\cite{bib:13} due to the characteristics of excellent energy resolution and imaging capability.

The PandaX-III experiment~\cite{bib:14},~\cite{bib:15},~\cite{bib:16} employs a high-pressure xenon gaseous TPC with large-area charge readout modules to search the NLDBD of \ce{^{136}Xe}.
Accurate calibration of the TPC is essential for understanding detector performance, which significantly impacts the accuracy of the experimental results, reliability, and interpretability.
Consequently, adopting an efficient, convenient, and comprehensive calibration method for large-area readout detectors is imperative.
Currently, calibration schemes utilizing radioactive sources~\cite{bib:17,bib:18} and cosmic rays~\cite{bib:19,bib:20} are widely used.
Radioactive sources such as \ce{^{55}Fe}, \ce{^{241}Am}, and \ce{^{232}Th} are commonly used as a point source, however, effectively utilizing such sources for large active area detectors is not ideal.
It is also difficult to mount and remove the source inside the TPC volume after the detector has been commissioned.
Cosmic rays with penetrating power and relatively uniform energy deposition per unit length allow us to characterize key performances of TPCs, such as the drift velocity and electron lifetime~\cite{bib:21}.
However, the continuous energy spectrum of cosmic rays causes non-negligible errors in calculating the gain and energy resolution.

Fast decaying gaseous sources, such as \ce{^{220}Rn}, \ce{^{83m}Kr}, and \ce{^{37}Ar}, can be directly mixed into the gas medium and provide an alternate method for the calibration of the gaseous TPC~\cite{bib:22,bib:23}.
\ce{^{37}Ar} decays to stable \ce{^{37}Cl} by electron capture with a half-life of 35 days, and the vacancy from one of the shells (K-, L-, or M-shell) is filled by an electron rearrangement accompanied by a cascade emission of x-rays and Auger-Meitner electrons. 
The overall energy deposition of \ce{^{37}Ar}, corresponding to the K-, L-, and M-shell electron binding energies, results in peaks at 2.82 keV, 0.27 keV, and 0.01 keV, respectively.
The first demonstration of a sub-keV electron recoil energy threshold in a dual-phase liquid argon time projection chamber is obtained with an \ce{^{37}Ar} source by observing the peaks in the energy spectrum at 2.82 keV and 0.27 keV~\cite{bib:24}.
The low-energy peaks of \ce{^{37}Ar} have also been used to extend the low-energy threshold of dual-phase xenon TPCs for dark matter detection experiments~\cite{bib:25},~\cite{bib:26},~\cite{bib:27}.

We present a comprehensive internal calibration method using an \ce{^{37}Ar} gaseous radioactive source for a gaseous TPC at different working pressures ranging from 0.3 bar to 10 bar, which represents a novel attempt to introduce a reference source into the TPC active volume.
An \ce{^{37}Ar} source has two advantages over a solid point source traditionally used for gaseous TPCs.
The unique combination of homogeneous distribution in the TPC volume, quasi-point source energy deposition of low-energy peaks, and gaseous TPC's event-by-event position reconstruction capability makes \ce{^{37}Ar} highly efficient in calibrating the detector response uniformity and electric field distortion. 
\ce{^{37}Ar} can be easily dissolved into and separated from the detector gas, which is convenient for real-time periodic calibration of the detector.

The gaseous TPC has an active volume of approximately 8~L and a maximum drift distance of 20.0~cm.
The detector can be operated up to 15 bar, and the working gas is an argon-(2.5\%)isobutane gas mixture in this work. 
The isobutane, used as a quencher in argon gas, improves the detector gain and resolution attributed to the Penning effect~\cite{bib:28}.
To ensure insulation and support for the detector, an electric field shaping cage (field cage, in short) made of an acrylic barrel with a diameter of 34.0~cm and a height of 20.0~cm is employed.
The field cage comprises copper rings, on which voltage divider resistors are carefully arranged to generate a uniform drift electric field in the drift volume.
One Micromegas module (Micro-Mesh Gaseous Structure)~\cite{bib:29} fabricated with the so-called thermal-bonding technology~\cite{bib:30,bib:31} in the University of Science and Technology of China is used for charge readout in the TPC. 
The Micromegas is a proven Micro Pattern Gas Detector (MPGD) option, consisting of a thin metallic grid (commonly called mesh) and an anode plane separated by insulating pillars. 
Both structures define a small avalanche gap (between 20 and 300 ${\rm \mu m}$), where ionization electrons generated in the drift volume are amplified.
This technology is widely used in particle and nuclear physics.
It has proven to have many advantages, such as its high granularity, good energy resolution, easy construction, gain stability, and sufficient radiopurity of the materials used to make these charge readouts.

In the internal calibration, the 2.82-keV peak from the \ce{^{37}Ar} source is used to calculate the gain and energy resolution of the detector under different pressures.
Correspondingly, the electron transmission and gain curves are presented with various amplification fields and drift fields.
The gain uniformity of the detector, mainly resulting from the homogeneity of the avalanche gap between the mesh and the anode of Micromegas~\cite{bib:32}, is calculated with uniformly distributed \ce{^{37}Ar} decay events and then used to improve the reconstructed charge and subsequently energy resolution of acquired events.
We have also quantified the drift field evolution with \ce{^{37}Ar} decay events, which impacts the effective area of the detector.
The corresponding results are corroborated with the simulation using the COMSOL Multiphysics software~\cite{bib:33}. 

This article first details the experimental setup in Section~\ref{sec.II}.
Then, the experimental procedure and data taking are described in Section~\ref{sec.III}. 
The results and discussion in terms of electron transmission, gain, energy resolution, gain uniformity, and drift field evolution with \ce{^{37}Ar} calibration method are explained in Section~\ref{sec.IV}.
Finally, the conclusions are presented in Section~\ref{sec.V}.

\section{Experimental setup}\label{sec.II}

\begin{figure}[tb]
    \includegraphics[width=\hsize]{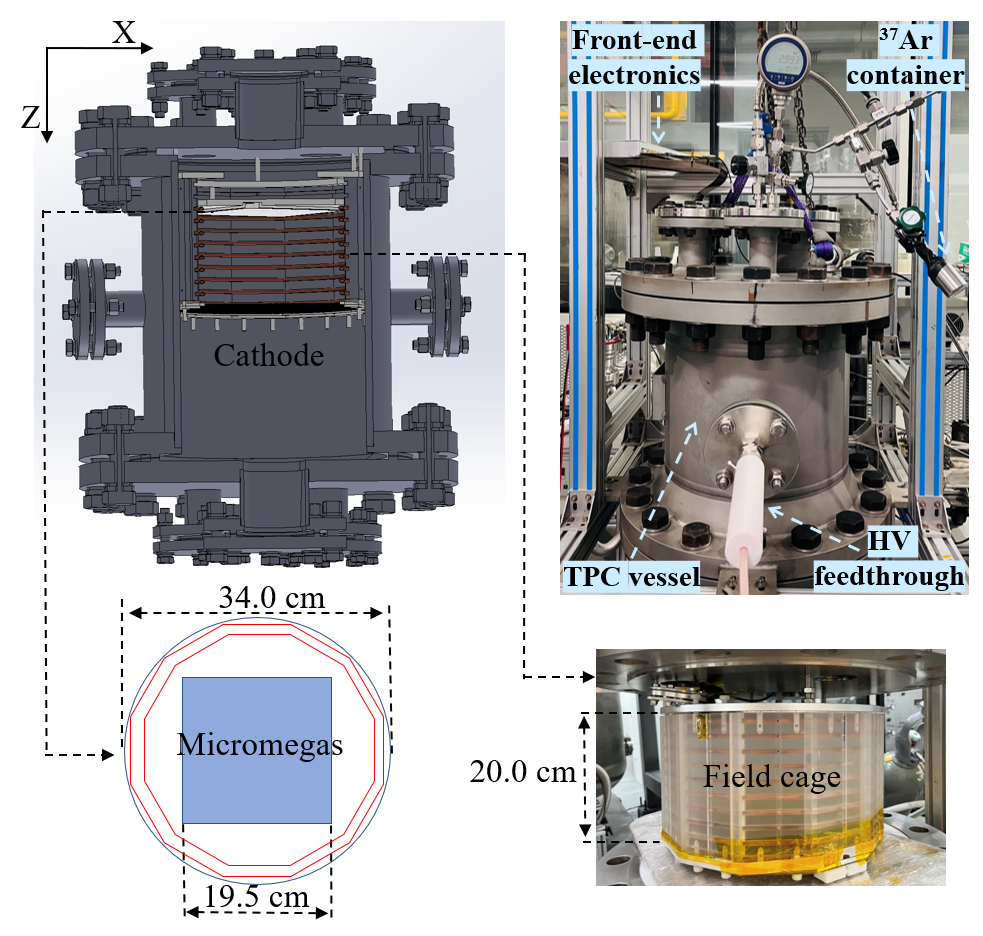}
    \caption{(Color online) Schematic diagrams (left-top) of the experimental setup, including the Micromegas readout plane (left-bottom), the cathode, and the field cage (right-bottom). The right-up panel displays a picture of the experimental setup.}
    \label{TPC_setup}
\end{figure}

The experimental setup is depicted in Fig.~\ref{TPC_setup}.
The core component is a gaseous TPC, which consists of a field cage with a height of 20.0 cm, a square-shaped readout plane with a side length of 19.5 cm on the top, and a cathode with a diameter of 34.0 cm on the bottom.
The active volume of the TPC is approximately 8~L, which defines the drift volume enclosed by the readout plane, the field cage, and the cathode.
The readout plane comprises one Micromegas module mounted on a circular aluminum holding plate, defined as the horizontal XY plane, with the drift field pointing towards the vertical Z direction.
The dodecagonal-prism-shaped field cage is suspended from the holding plate of the readout plane.
It comprises an acrylic barrel that provides a mechanical structure, holding ten copper rings that serve as electrodes, each with a thickness of 6 mm and a spacing of about 20 mm.
Ten 1 ${\rm G \Omega}$ resistors are placed between adjacent sets of copper rings for degraded electric potentials.
The cathode is simply an aluminum mesh connected to the bottom ring of the field cage.
Negative high voltage (HV) is applied to the aluminum cathode through a feedthrough on the side of the TPC pressure vessel. 
In this configuration, electric field lines point from top to bottom, and ionization electrons drift upward in the field cage.

\begin{figure}[tb]
    \includegraphics[width=\hsize]{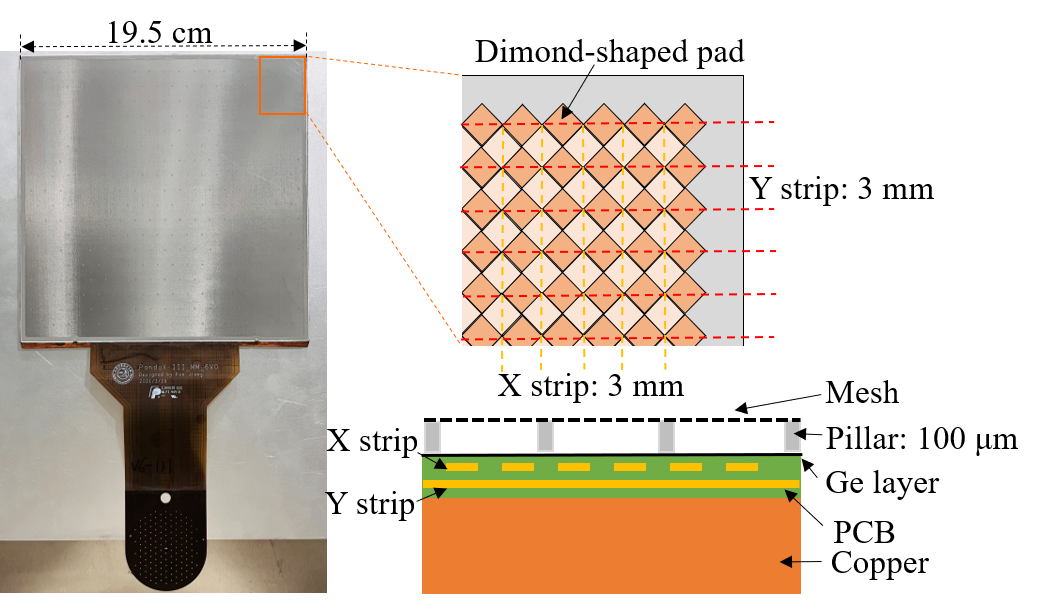}
    \caption{(Color online) The left picture presents a newly designed thermal-bonding Micromegas. Schematic diagrams on the right show the readout geometry and layer structure of Micromegas. }
    \label{Micromegas_setup}
\end{figure}

Fig.~\ref{Micromegas_setup} shows the Micromegas module fabricated with the thermal-bonding method, which makes the amplification structure by pressing a stainless steel (SS) mesh directly against insulating pillars on the readout printed circuit board (PCB) anode using a hot rolling machine~\cite{bib:30}.
The readout PCB is attached to a high-purity, oxygen-free copper substrate to provide a rigid support for resisting the tension of the SS mesh.
A germanium (Ge) layer is deposited on the readout PCB as a resistive anode to suppress discharges and improve the performance of the detector~\cite{bib:34}.
The insulating pillars with thermal-bonding epoxy are placed on the Ge layer to support the SS mesh and create an avalanche gap of 100 ${\rm \mu m}$.
The SS mesh has 325 lines per inch with a wire diameter of 19~${\rm \mu m}$.
Negative HV is applied to the mesh through the traces on the flexible tail of Micromegas.
The Micromegas readout module has an active area of ${19.5~\text{cm} \times 19.5~\text{cm}}$ with a tightly arranged group of diamond-shaped readout pads, which are interconnected into strips, as shown in Fig~\ref{Micromegas_setup}. 
The Micromegas has 64 readout strips in each direction (X and Y) embedded in the inner layer of the readout PCB anode and then connected to the electronics via the flexible tail for signal readout.
The strip pitch is 3 mm.
The strip readout scheme is chosen to reduce the number of readout channels in the readout PCB anode, and the energy deposition on the Micromegas is shared between X and Y strips.

Thermal-bonding technology has the advantages of low-cost, straightforward production procedures and convenient gap size adjustment through pillar replacement while maintaining excellent performances of the manufactured detectors.
The avalanche gap size depends on the pillar thickness, and its uniformity across the active area is critical for achieving high energy resolution and stable operation.
The surface smoothness of the Micromegas anode plate and the conditions such as temperature and pressure in the thermal-bonding process can affect the size of the avalanche gap as well.

When an incident particle travels in the active volume of the TPC, it deposits energy by ionizing the gas atoms along its trajectory. 
With a negative high voltage applied to the cathode, the primary ionization electrons drift along the electric field and get collected by the Micromegas strips in the X and Y directions after avalanche amplification.
The electronics and Data Acquisition system (DAQ) reads out triggered signals from Micromegas strips and subsequently digitizes the resultant signals and writes the data to disk.
The strip signals are digitized by the ASIC Support and Analog Digital conversion (ASAD) commercial front-end board and a Concentration Board (CoBo) back-end board~\cite{bib:35}, based on the ASIC for Generic Electronic system for TPCs (AGET) chip~\cite{bib:36}.
The AGET chips are widely used in gaseous TPCs for particle and nuclear physics experiments.
They have sampling frequencies of up to 100 MHz, dynamic ranges from 120~fC to 10~pC, and peaking times from 50~ns to 1~${\rm \mu s}$. 
The wide dynamic ranges and adjustable sampling frequencies enable us to flexibly obtain event data with various optional configurations to satisfy different experimental requirements.
A Field Programmable Gate Array (FPGA) on the ASAD board controls the AGET chips and sends the packets of collected data to the CoBo card.
A CoBo card reads up to four ASAD boards, and an ASAD board hosts four AGET chips, each of which can process 64 strip channels from the detector input.
Therefore, in the configuration of this experiment, two AGET chips are activated for the Micromegas module with 128 strip channels, half of which are in the X direction and the other half in the Y direction.
The DAQ system we use is shipped with the ASAD/CoBo combination. It reads and writes configuration parameters of ASAD/CoBo, manages the data-taking process, monitors data quality, and makes a subset of acquired waveforms available for inspection.

The field cage is suspended from the top flange of the SS pressure vessel described earlier, which can be operated at pressures up to 15 bar.
The pressure vessel consists of a cylindrical main body with a thickness of 1.8~cm, in conjunction with top and bottom flanges, both of which are 3.5~cm thick.
The inner diameter of the cylindrical main body is about 38.8~cm and the height is 47.0~cm, creating an inner volume of approximately 60~L. 
The inner volume of the SS vessel is much larger than the TPC drift volume to avoid electrical discharges from the field cage electrodes to the ground.
Two DN-50 ports are located in the middle of the cylindrical main body for high-voltage feedthrough of the cathode and vacuum pumping.
The pump port is connected to a dry fore-pump (PTS300 from Agilent) and a turbo pump (HiPace300 from Pfeiffer), and a vacuum of better than ${10^{-5}}$~mbar can be achieved.
The pump outlet is connected to a flexible pipeline that extends outside the laboratory, facilitating proper ventilation to ensure the safety of personnel. 
A leakage check is performed before filling the gas source, and a Geiger counter is positioned around the detector to monitor environmental radiation levels.
The flat top and bottom flanges are fastened onto the cylindrical main body by 16 M33 bolts.
Three ports on the top and bottom flanges are utilized for Micromegas signals and HV bias, gas inlet and outlet, and as backups, respectively.
Micromegas signals from the readout plane are transmitted through the DN-65 port of the top flange via a custom-made Kapton extension cable feedthrough.

\begin{figure}[tbp]
    \includegraphics[width=\hsize]{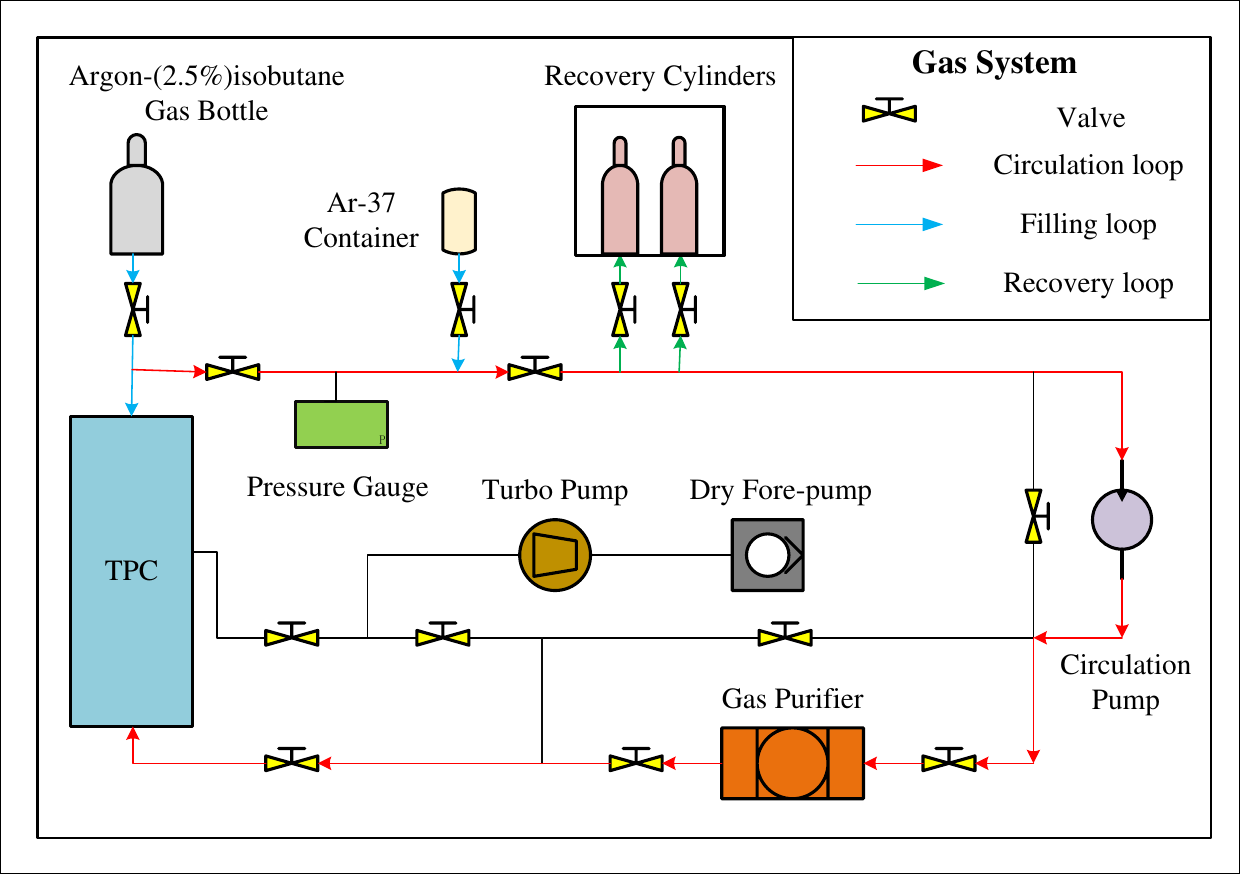}
    \caption{(Color online) Schematic diagram of the gas system of the TPC.}
    \label{gasSystem}
\end{figure}

The gas system of the TPC is shown in Fig.~\ref{gasSystem}.
It is connected to the gas inlet and outlet of the pressure vessel, performing gas filling, circulation, purification, and recovery for the experiment.
Due to the outgassing of the detector materials like acrylic, the quality of the gas mixture in the vessel deteriorates over time. To mitigate this, continuous purification with a circulation loop is necessary. 
A room-temperature gas purifier~\cite{bib:37} installed in the gas system can effectively eliminate impurities such as ${\rm H_{2}O}$, ${\rm O_{2}}$, CO, and ${\rm CO_{2}}$ from gases with the assistance of a magnetically driven circulation pump.
Gas recovery can be performed in the line of the gas system with two 500~ml recovery cylinders.
When immersed in a liquid nitrogen dewar during operation, the recovery cylinders function as an absorption pump and liquify argon for storage.
All the pipes are connected using Swagelok VCR metal gasket face seal fittings, with designed working pressures of up to 15 bar. 

A small \ce{^{37}Ar} SS container (25~ml volume, 0.1~bar pressure) and a pre-mixed argon-(2.5${\rm \%}$)isobutane gas bottle (40~L volume, 40~bar pressure) are connected onto the gas system for gas filling. 
The isotope \ce{^{37}Ar} is produced by irradiation of 99.9355\%-enriched \ce{^{36}Ar} gas with thermal neutrons: ${\ce{^{36}Ar}(n,\gamma )\ce{^{37}Ar}}$. 
The enriched \ce{^{36}Ar} contained in a quartz glass ampoule is used for irradiation with a capture cross-section of about 5~barn.
The enrichment enhances the production of \ce{^{37}Ar} and reduces the production of undesired isotopes.
After irradiation, the quartz glass ampoule is shattered so that the activated gas is transferred to a small SS container and depressurized to 0.1 bar.
An initial \ce{^{37}Ar} activity of about 100~Bq per SS container is expected according to the neutron flux, the neutron capture cross-section of \ce{^{36}Ar}, and the irradiation time. 

\section{Experimental procedure and data taking}\label{sec.III}

The TPC pressure vessel is pumped below ${\rm 1 \times 10^{-5}}$ mbar before injecting the activated \ce{^{37}Ar} gas (activity: 100~Bq) from a 25~ml SS container.
The detector gas is then injected into the vessel to 1 bar.
Two recovery cylinders in a liquid nitrogen dewar are connected to the vessel to reduce the pressure to 0.3 bar by liquifying the gas mixture.
After the low-pressure measurement is completed, 1 bar of gas in the vessel is recovered from two recovery cylinders through the gas circulation system, and then the gas pressure is increased to 10 bar by injecting more pre-mixed gas for high-pressure measurements.

The \ce{^{37}Ar} K-shell decay events deposit energy by ionizing gas atoms and the ionization electrons are collected by the readout strips.
The strip signals are digitized by the ASAD-CoBo electronics with a record length of 512 sampling points and a sampling frequency of 10~MHz (time window: 51.2~${\rm \mu s}$).
The horizontal position of an event is determined by triggered strips in the XY readout plane, while the absolute Z position and start time (${t_{0}}$) require scintillation light information.
Relative Z positions are instead inferred from the timing of signal waveforms.
Therefore, a complete three-dimensional image of an event is recorded.

\begin{figure}[tbp]
    \subfigure{
    \label{source_pulse}
    \includegraphics[width=0.8\hsize,trim=0 0 0 0,clip]{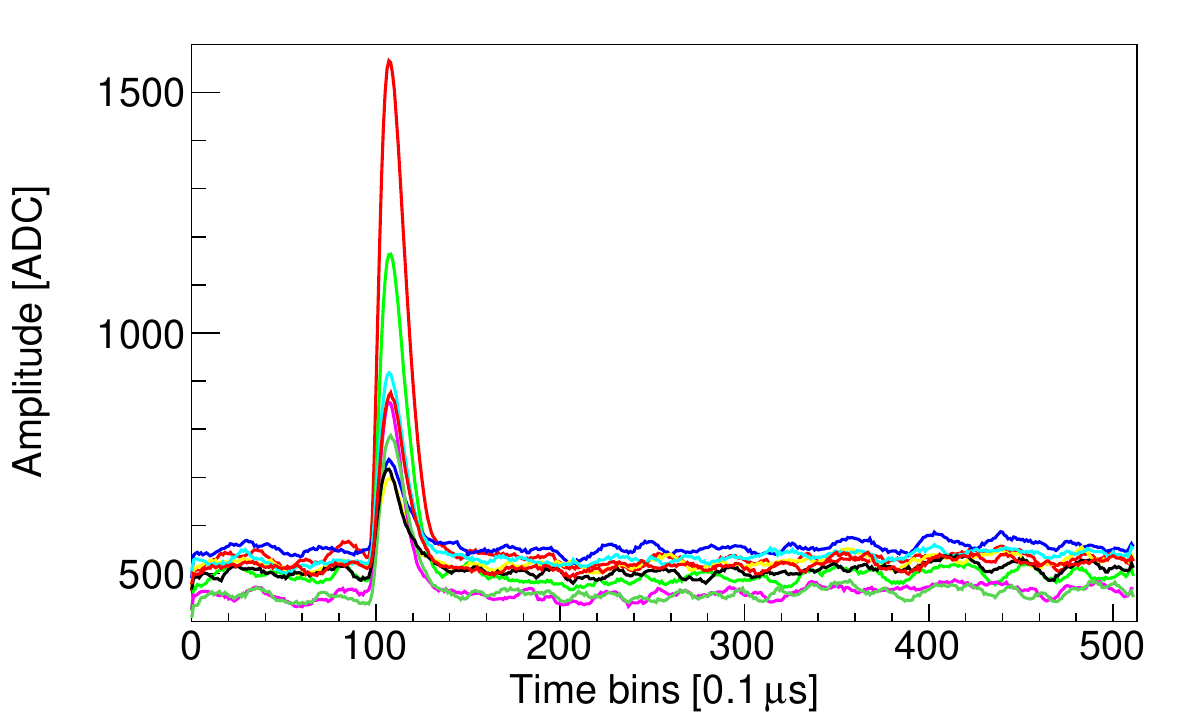}
 }
    \subfigure{
    \label{source_imagexz}
    \includegraphics[width=0.46\hsize,trim=20 0 20 0,clip]{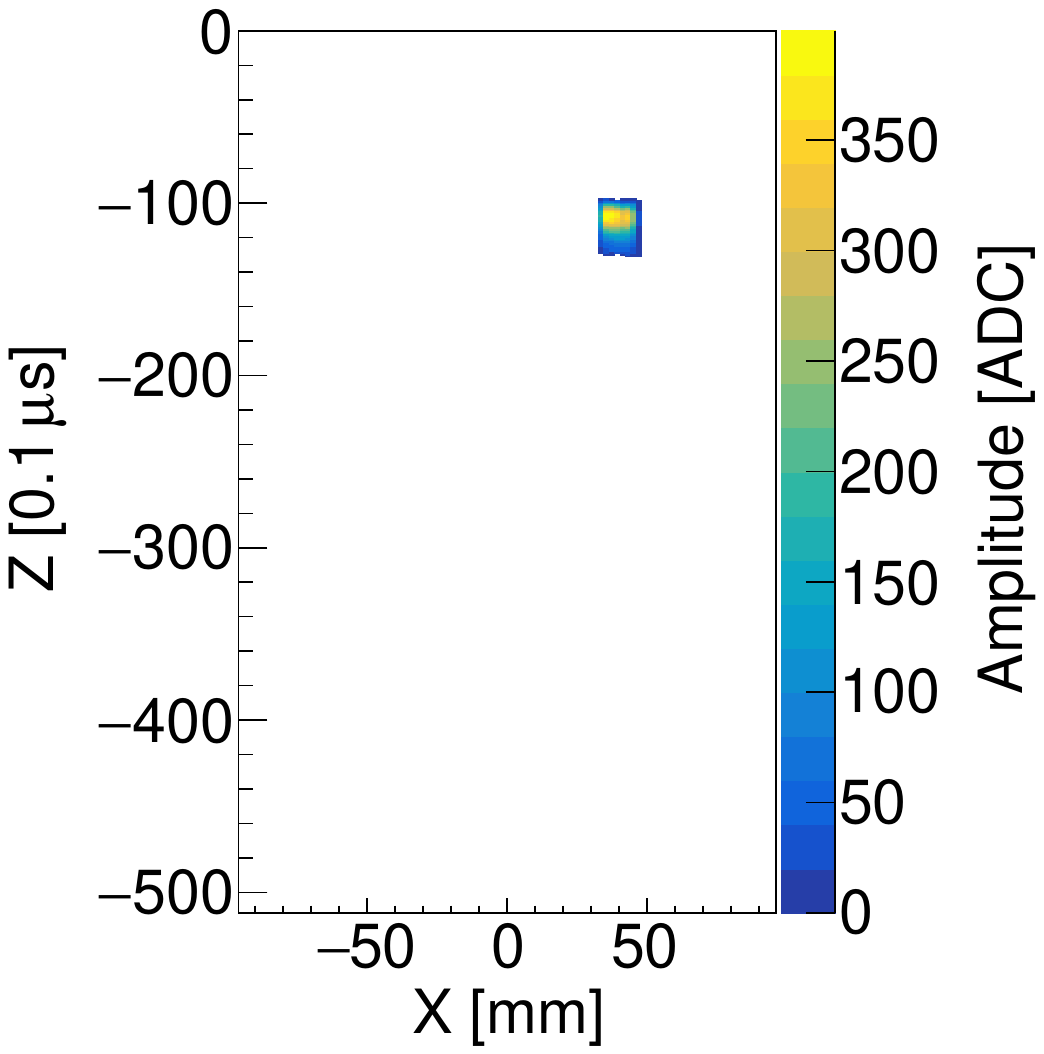}
 }
    \subfigure{
    \label{source_imageyz}
    \includegraphics[width=0.46\hsize,trim=20 0 20 0,clip]{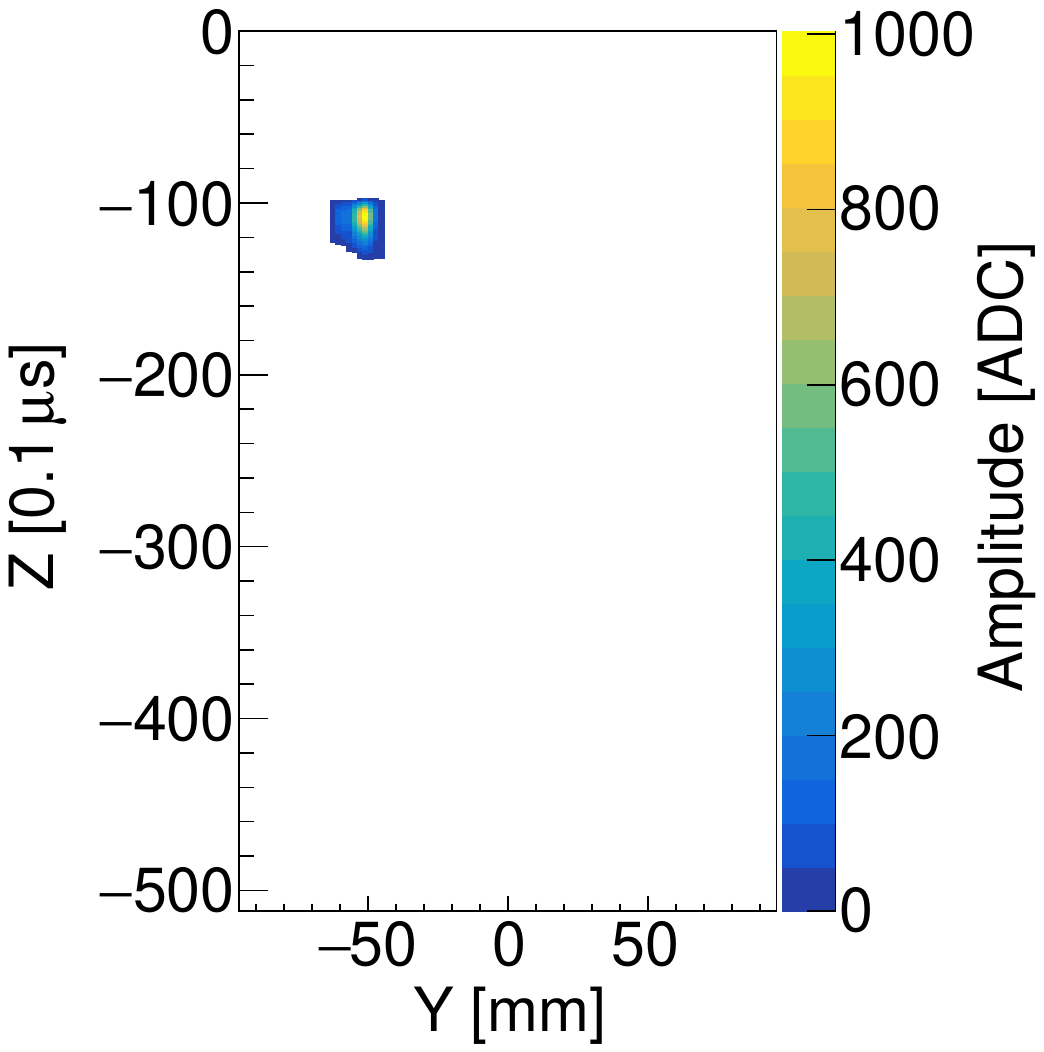}
 }
    \caption{(Color online) (top) Waveforms of an \ce{^{37}Ar} K-shell decay event, where one color indicates one strip. (bottom) Images of an \ce{^{37}Ar} K-shell decay event on the XZ and YZ planes.}
    \label{sourceSignals}
\end{figure}

An example of \ce{^{37}Ar} decay events, including their waveforms and projected images, is shown in Fig.~\ref{sourceSignals}.
Data are acquired with our electronics system in a 1 bar detector gas. 
In the DAQ configuration, 112 sampling points (11.2~${\rm \mu s}$) before and 400 sampling points (40.0~${\rm \mu s}$)after the trigger are recorded.
The projected images show small-sized electron clusters generated by low-energy source events. 
The average triggered X and Y strips are approximately 6 at 1 bar.
The colors in the image represent the amplitude of the waveform for each event, which corresponds to the collected charge.
Data unpacking and analysis are performed in the REST (Rare Event Searches Toolkit for Physics)~\cite{bib:38} framework, which is a ROOT-based solution for event-oriented data analysis and Monte Carlo simulation specifically designed for gaseous TPCs.

\begin{figure}[htp]
    \includegraphics[width=\hsize]{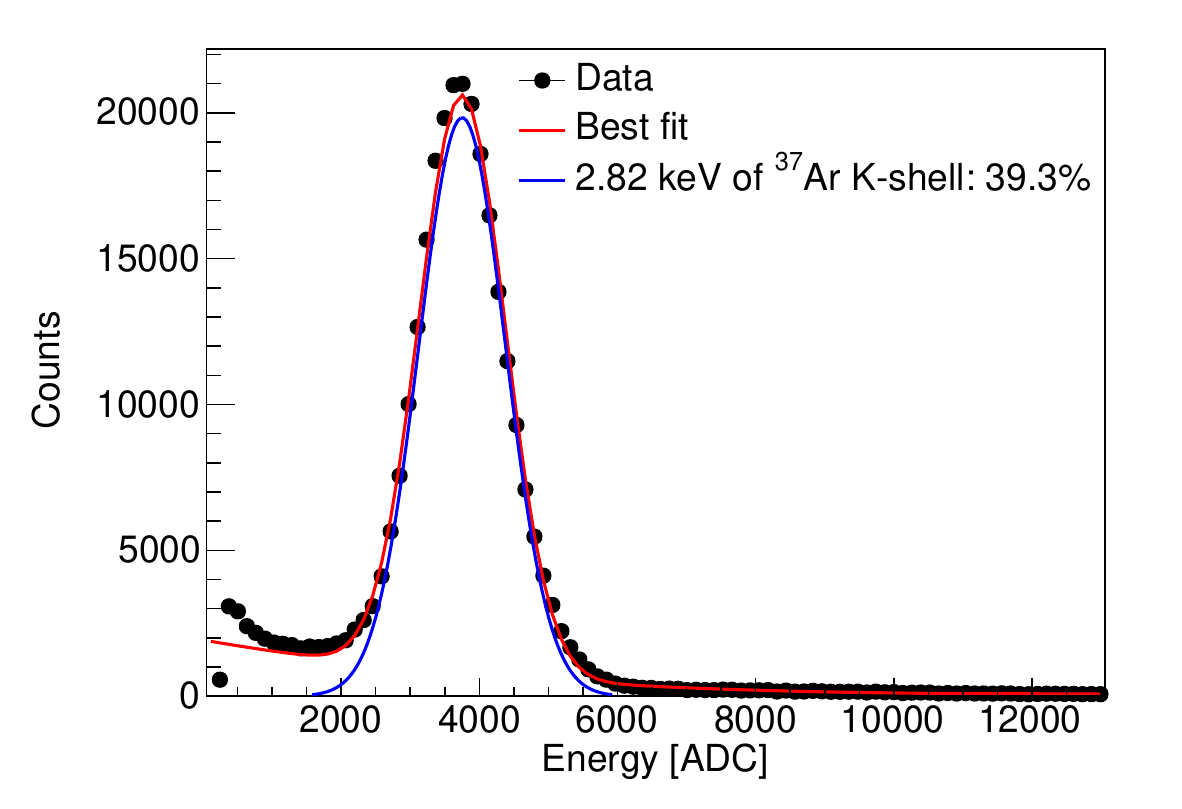}
    \caption{(Color online) A typical energy spectrum of the \ce{^{37}Ar} K-shell decay events in 1 bar argon-(2.5${\rm \%}$)isobutane gas mixtures.}
    \label{spe_1bar}
\end{figure}

The energy (ADC unit) of an event is determined by summing the waveform amplitudes of over-threshold signals on strips.
Fig.~\ref{spe_1bar} shows a typical energy spectrum of \ce{^{37}Ar} decay events at 1 bar, where the 2.82-keV peak is fitted with a Gaussian plus a polynomial function to eliminate the linear background.
The energy resolution of the detector is expressed as the full-width-at-half-maximum (FWHM) of the Gaussian function.
An energy resolution of 39.3\% at 2.82 keV is obtained. 
The peak is used to calculate the detector's gain, defined as the ratio of the total charges of primary ionization electrons after and before the detector amplification.
The average ionization energy is taken as 26.2~eV in argon-(2.5${\rm \%}$)isobutane gas mixtures.
The detector gain is approximately $6.5\times 10^3$ for the peak shown in Fig.~\ref{spe_1bar}, which corresponds to an amplification field of 37~kV/cm and a reduced drift field of 230~V/cm/bar.
Using the acquired \ce{^{37}Ar} data, a wide range of amplification and reduced drift fields has been scanned to study the TPC performance, including electron transmission, gain, energy resolution, gain uniformity, and drift field evolution. 

\section{Results and discussion}\label{sec.IV}

\subsection{Electron transmission, gain and energy resolution}

\begin{figure*}[htb]
    \centering
    \subfigure[]{
    \label{transmissionVsRatio}
    \includegraphics[width=0.48\hsize]{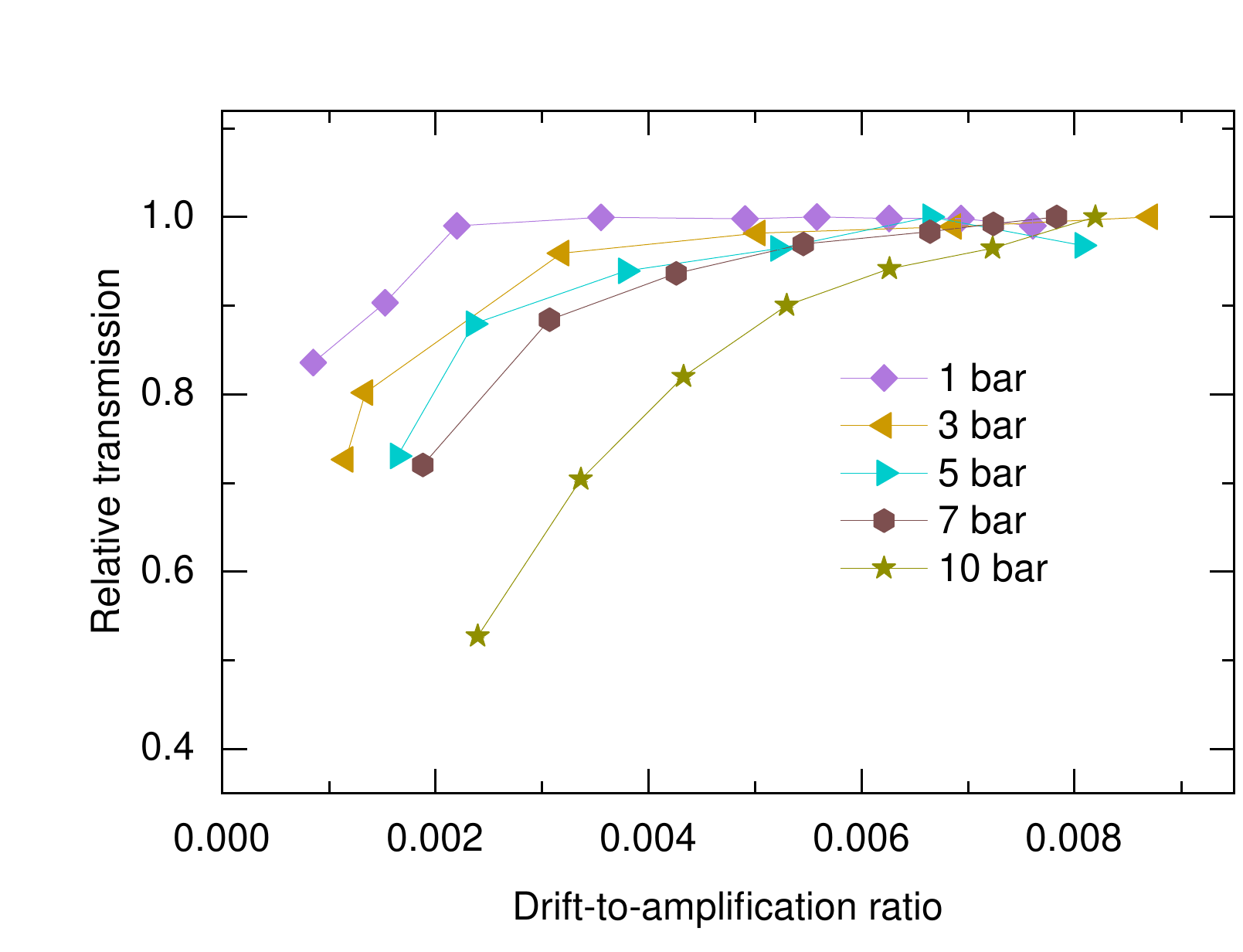}
 }
    \subfigure[]{
    \label{resolutionVsDrift}
    \includegraphics[width=0.48\hsize]{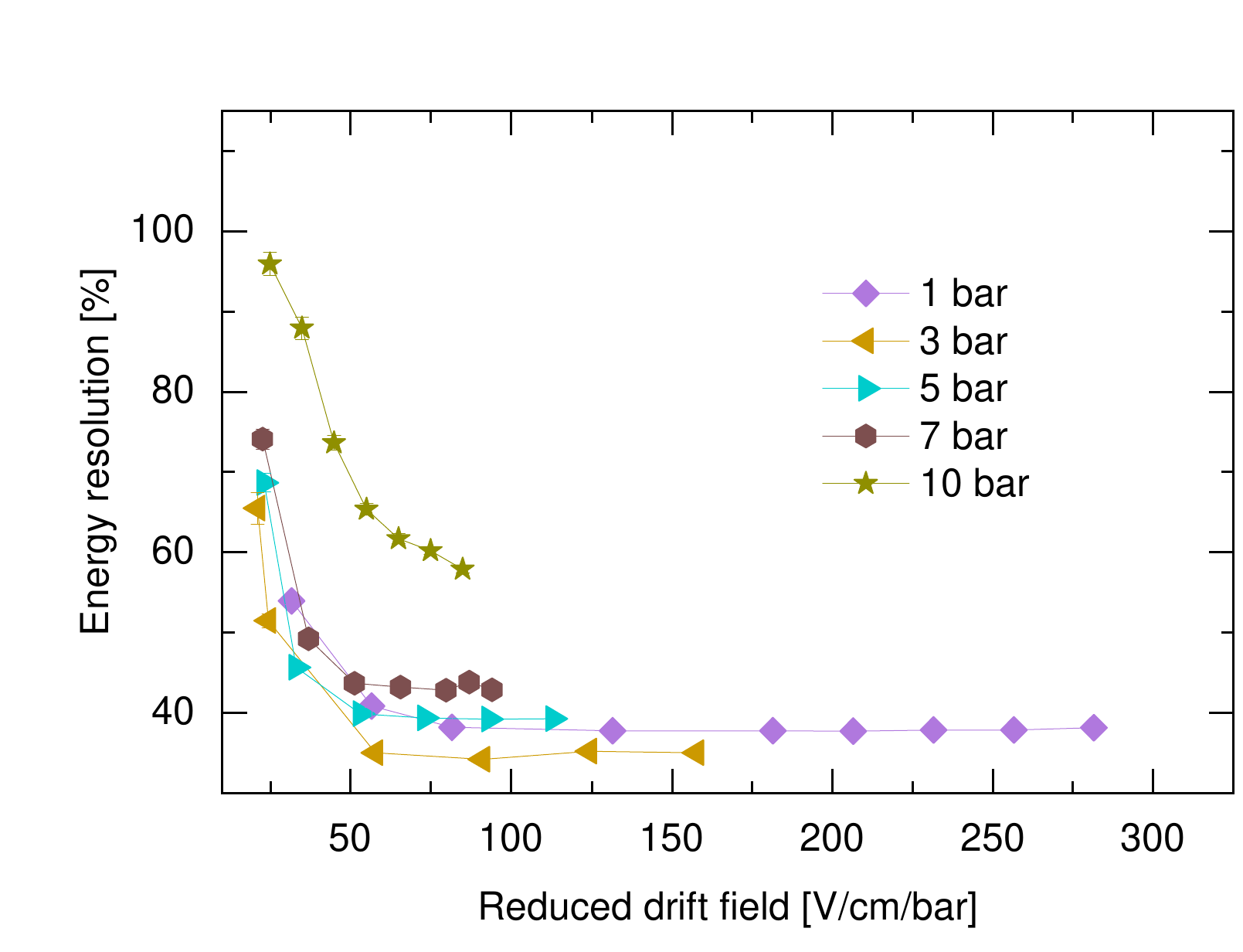}
 }
    \caption{(Color online) (a) Dependence of the relative transmission on the drift-to-amplification field ratio. (b) Dependence of the energy resolution on the reduced drift field.}
    \label{driftScan}
\end{figure*}

Electron transmission is the probability of ionization electrons passing through the Micromegas mesh holes from the drift volume to the avalanche gap.
The relative transmission is defined as the gain normalized to the maximum value of each series of measurements.
The drift voltage is varied at a fixed amplification voltage to obtain the dependence of the electron transmission on the drift-to-amplification field ratio from 1~bar to 10~bar, as shown in Fig.~\ref{transmissionVsRatio}.
At low drift fields, the electron transmission is reduced by electron attachment and recombination of ionization electrons generated in the drift volume.
At high drift fields, the plateau of electron transmission appears, indicating that the electron collection efficiency of the mesh is the highest and the attachment and recombination contributions are negligible.
We selected drift voltages in the plateau to ensure the most optimized electron transparency in the following tests.
Accordingly, the dependence of the energy resolution on the reduced drift field (Fig.~\ref{resolutionVsDrift}) shows its best value at the plateau for which the electron transmission is optimal.
However, it is more difficult to reach the plateau of electron transmission under high pressures, and a stronger reduced drift field is required.
Especially at a pressure of 10~bar, the energy resolution deteriorates significantly compared to low-pressure measurements.

\begin{figure*}[htb]
    \centering
    \subfigure[]{
    \label{gainVsAmplification}
    \includegraphics[width=0.48\hsize]{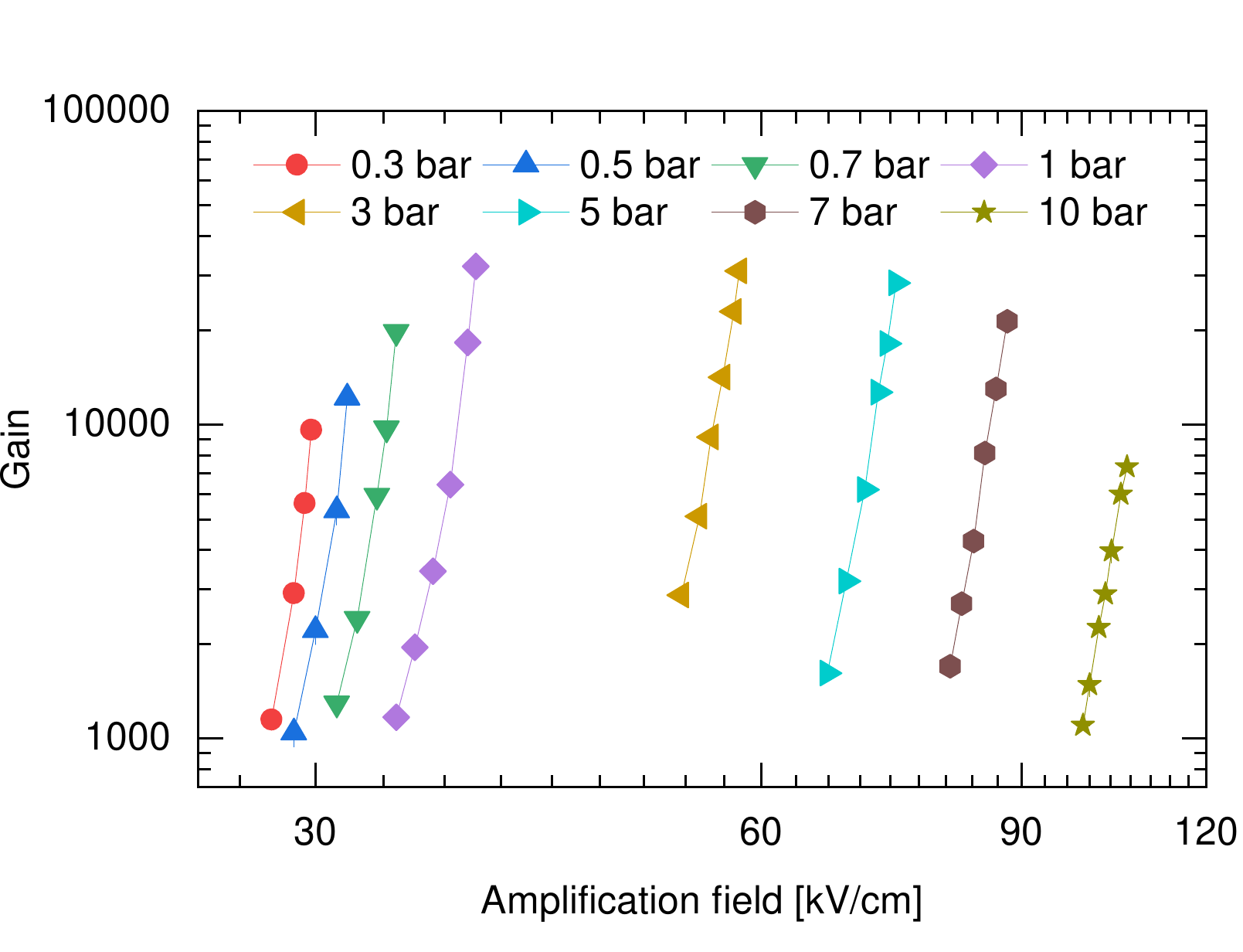}
 }
    \subfigure[]{
    \label{resolutionVsGain}
    \includegraphics[width=0.48\hsize]{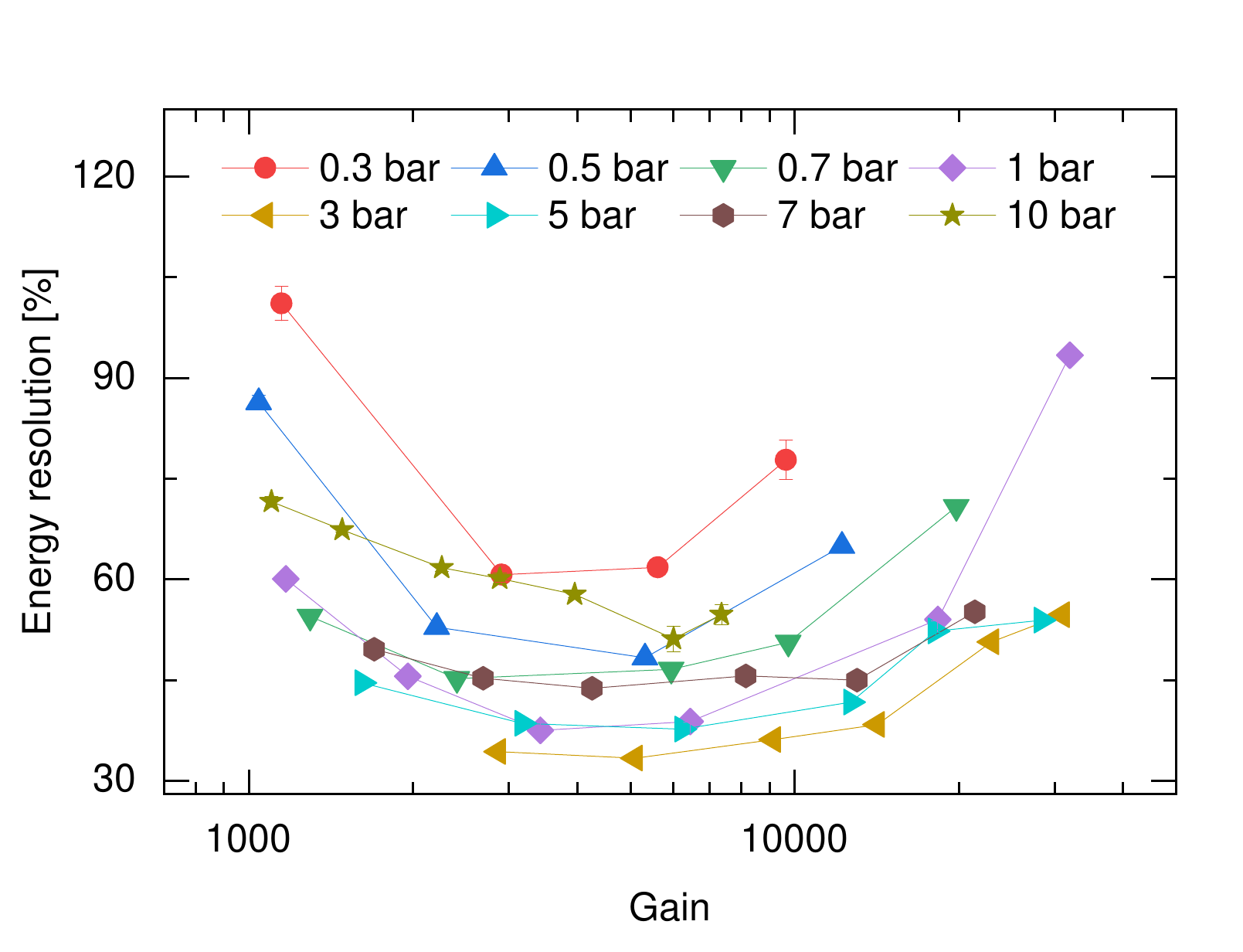}
 }
    \caption{(Color online) (a) Dependence of the gain on the amplification field. (b) Dependence of the energy resolution on the gain.}
    \label{meshScan}
\end{figure*}

The amplification voltage is varied to study the detector gain with the optimal drift voltages.
The gain curves in Fig.~\ref{gainVsAmplification} show a linear behavior with the amplification field on the log-log plot and achieve maximum attainable gains of up to around ten thousand at each pressure.
It demonstrates the excellent performance and adaptability of the detector under different pressures.
Among them, a maximum gain of $3.2\times 10^4$ is obtained with an amplification field of 38.5~kV/cm at 1~bar.
The maximum attainable gain reaches its highest value at atmospheric pressure. 
However, due to the Micromegas sparking, achieving a higher maximum attainable gain at very low (0.3 bar) and high (10 bar) pressures is difficult.
The energy resolution evolving with the amplification field, more specifically with the gain, is shown in Fig.~\ref{resolutionVsGain}.
The figure shows its dependence on the absolute gain for all pressures, and the statistical error is derived from the errors in the fitting parameters of energy spectra.
The resolution degrades at low gains where the signal is comparable to the electronic noise. 
At high gains, the increase in avalanche fluctuations also leads to a degradation of the resolution~\cite{bib:39}.
The energy resolution shows its best value at a gain of about $5\times 10^3$ under different pressures.
Specifically, the energy resolution of 33.4\% at 3~bar is slightly better than those at other pressures.
At lower pressure, the ionized electrons are distributed to more readout strips due to diffusion, which depends on the absolute Z position.
The energy resolution suffers from a convolution of cluster-dependence on Z and charge loss due to trigger threshold. 
At higher pressure, the gain non-uniformity increases significantly, which deteriorates the energy resolutions as well.  

\subsection{Gain uniformity}

The gain uniformity is a crucial property of the detector as it directly affects the detector's intrinsic energy resolution.
Thanks to the uniformly distributed \ce{^{37}Ar} source and its low-energy electron clusters, the gain of the entire sensitive area of the detector can be calibrated quickly and effectively
The hit position of an event in the XY readout plane is generated from the average triggered-strip position using the charge-weighted method.
The hit position indicates the central projection position of the electron cluster of an event.
The sensitive area is divided into 64${\rm \times}$64 regions in the XY readout plane, and the energy spectrum in each region is reconstructed and fitted following the same procedure as shown in Fig.~\ref{spe_1bar}.
Thus, the gain map and the energy resolution map of Micromegas can be obtained with the energy spectrum of \ce{^{37}Ar} in each region.

\begin{figure*}[htbp]
    \centering
    \hspace{2.5em}
    \subfigure[]{
    \label{subfig:uni_peak}
    \includegraphics[width=0.36\hsize]{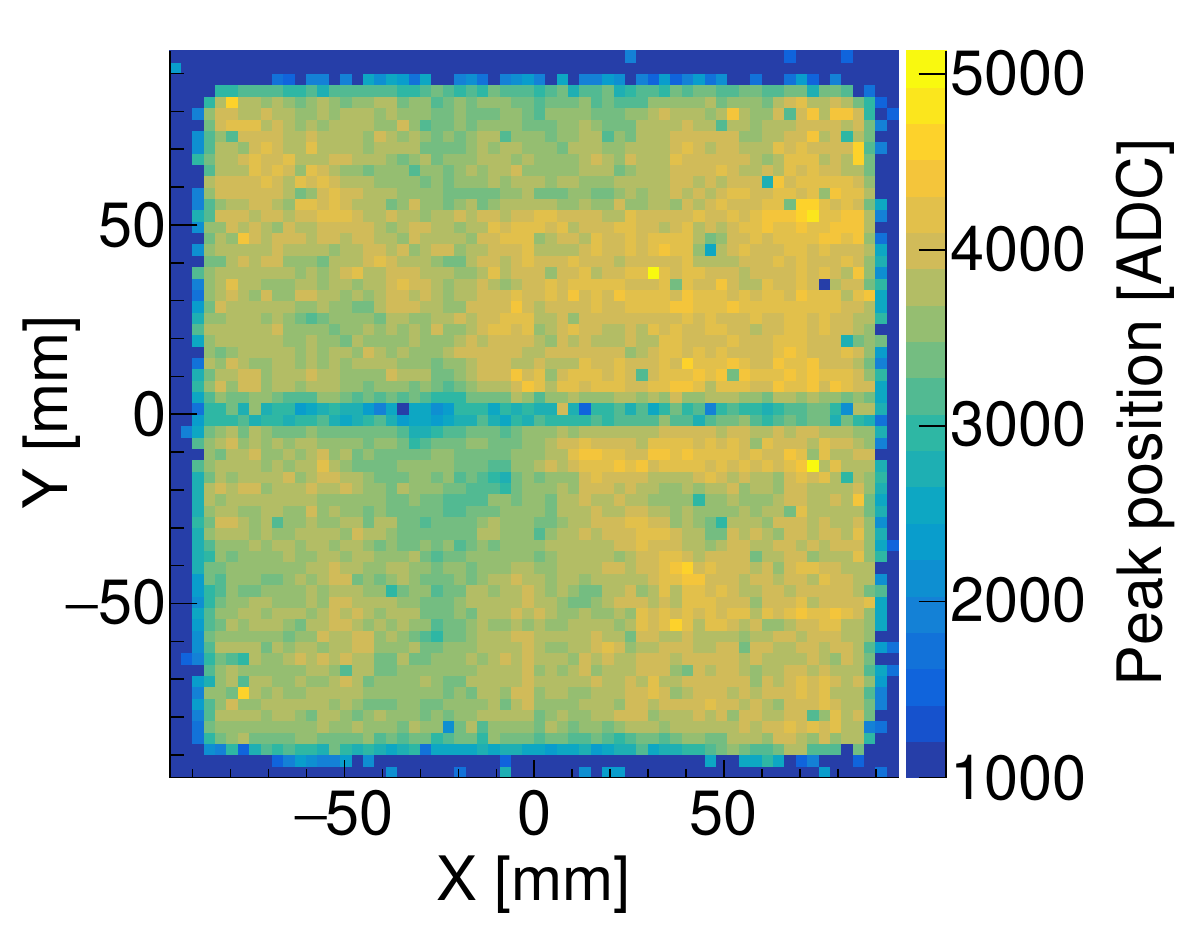}
 }
    \subfigure[]{
    \label{subfig:uni_resolution}
    \includegraphics[width=0.36\hsize]{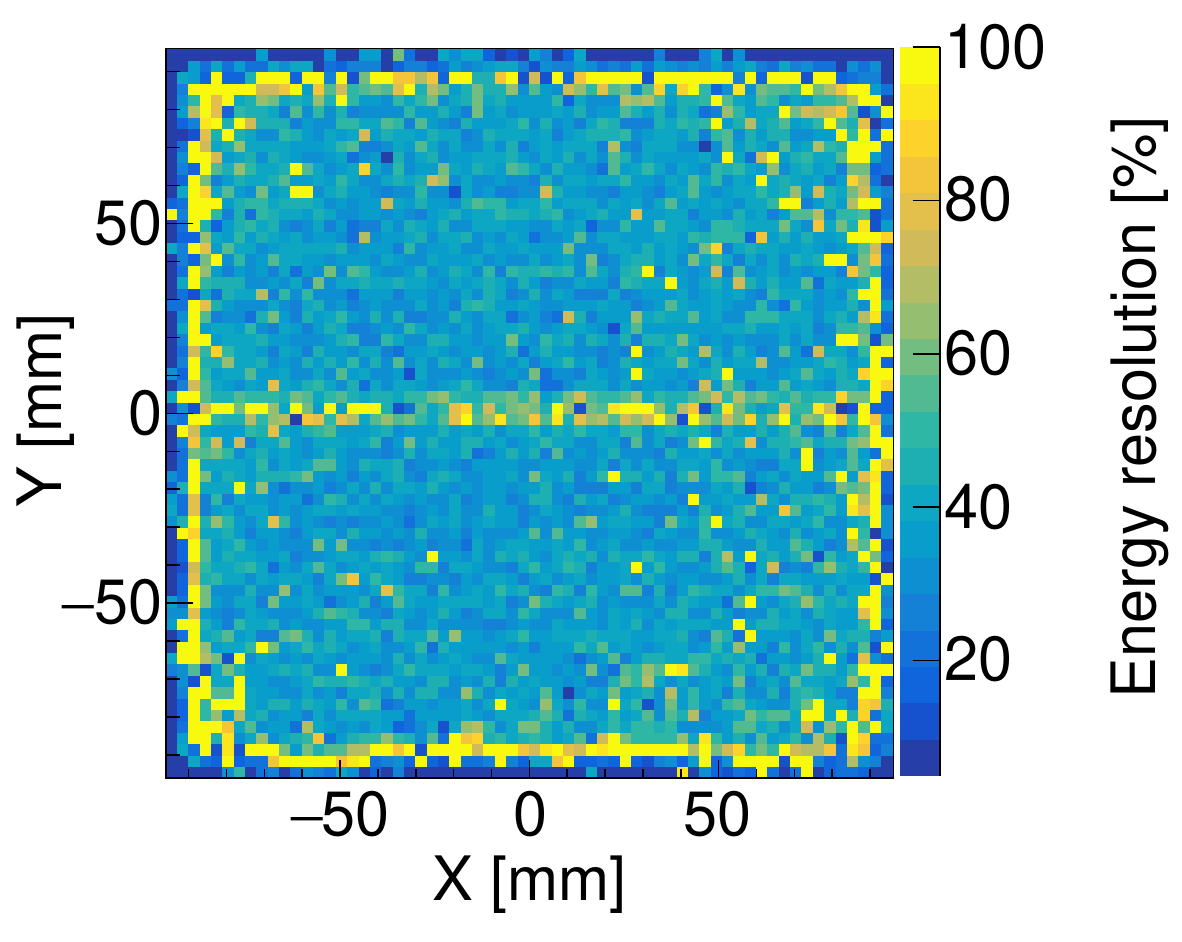}
 }
    \subfigure[]{
    \label{subfig:uni_peakRMS}
    \includegraphics[width=0.35\hsize]{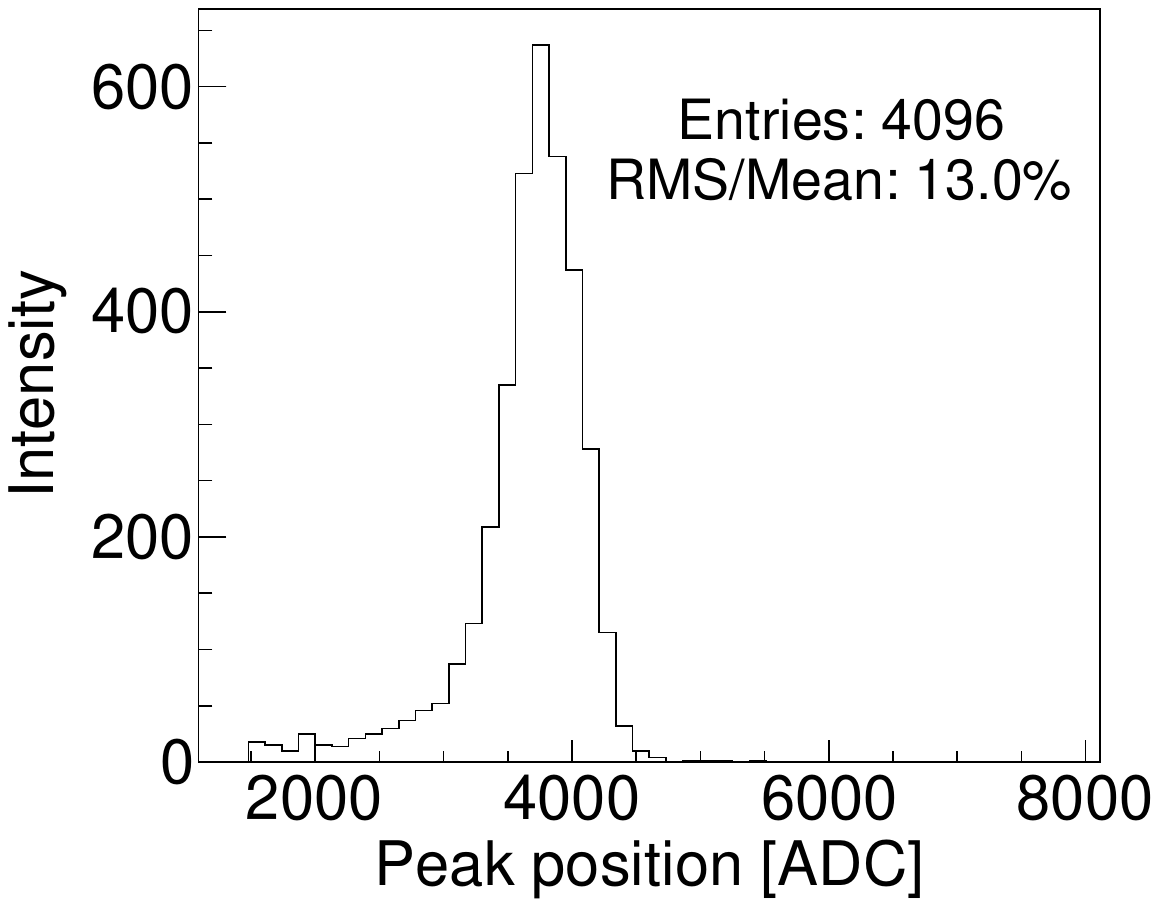}
 }
    \subfigure[]{
    \label{subfig:uni_resolutionRMS}
    \includegraphics[width=0.35\hsize]{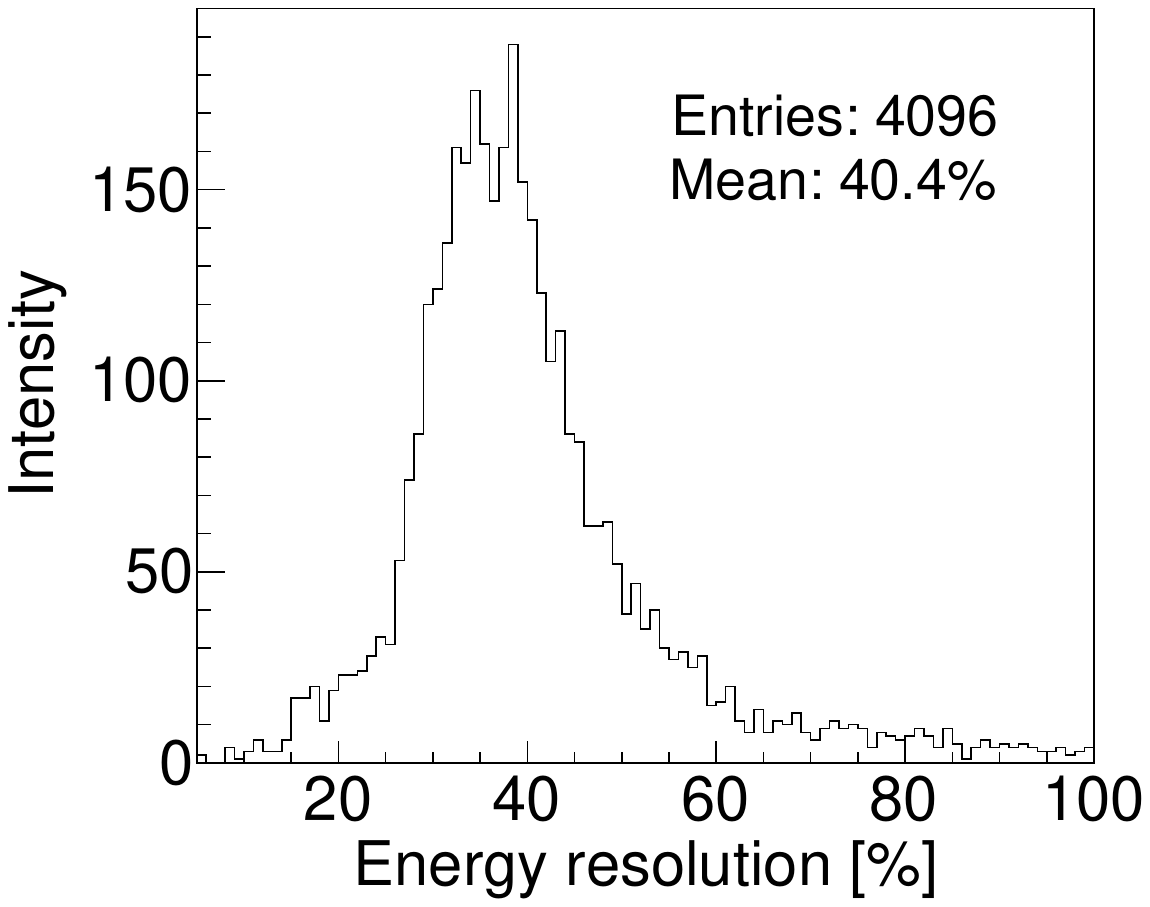}
 }
    \caption{(Color online) Typical two-dimensional map of the peak position (a) with the corresponding one-dimensional distribution over the 64${\rm \times}$64 regions (c). Typical two-dimensional map of the energy resolution at 2.82 keV (b) with the corresponding one-dimensional distribution over the 64${\rm \times}$64 regions (d).}
    \label{uni_map}
\end{figure*}

Fig.~\ref{uni_map} shows a typical distribution of the mean value (peak position) and standard deviation (energy resolution) of the fit function of energy spectra in each region with an amplification field of 37~kV/cm at 1 bar.
The gain non-uniformity is 13.0\% in this example, defined as the root mean square (RMS) of the peak distribution divided by their mean value.
In the central Y and detector boundary regions, the peak position is below 2000~ADC, and the energy resolution is above~80.0\%. 
However, the average peak position in other regions is about~3600, and the average energy resolution is about~40.4\%.
The low peak position of the central channel in the Y direction is due to a broken electronics channel. 
The Micromegas border is less effective in charge collection due to the electric field distortion, resulting in an extreme degradation of the peak position and energy resolution in these regions.
Moreover, for electron clusters of source events with average triggered strips of about six, falling at the edge of Micromegas also leads to incomplete charge collection of events.
Thus, the dead electronics channels and low-count channels in the Micromegas border are excluded from the non-uniformity calculation.
The obtained gain map, i.e. the peak position map, intuitively presents the gap uniformity of Micromegas and can be utilized to assess the detector's performance.
According to~\cite{bib:30}, the surface smoothness of the readout PCB anode can affect the gap uniformity of Micromegas.
Polishing and leveling the copper substrate and the PCB anode can enhance the gain uniformity from 13.9\% (13.2\%) to 11.2\%.

\begin{figure}[!htbp]
    \includegraphics[width=\hsize]{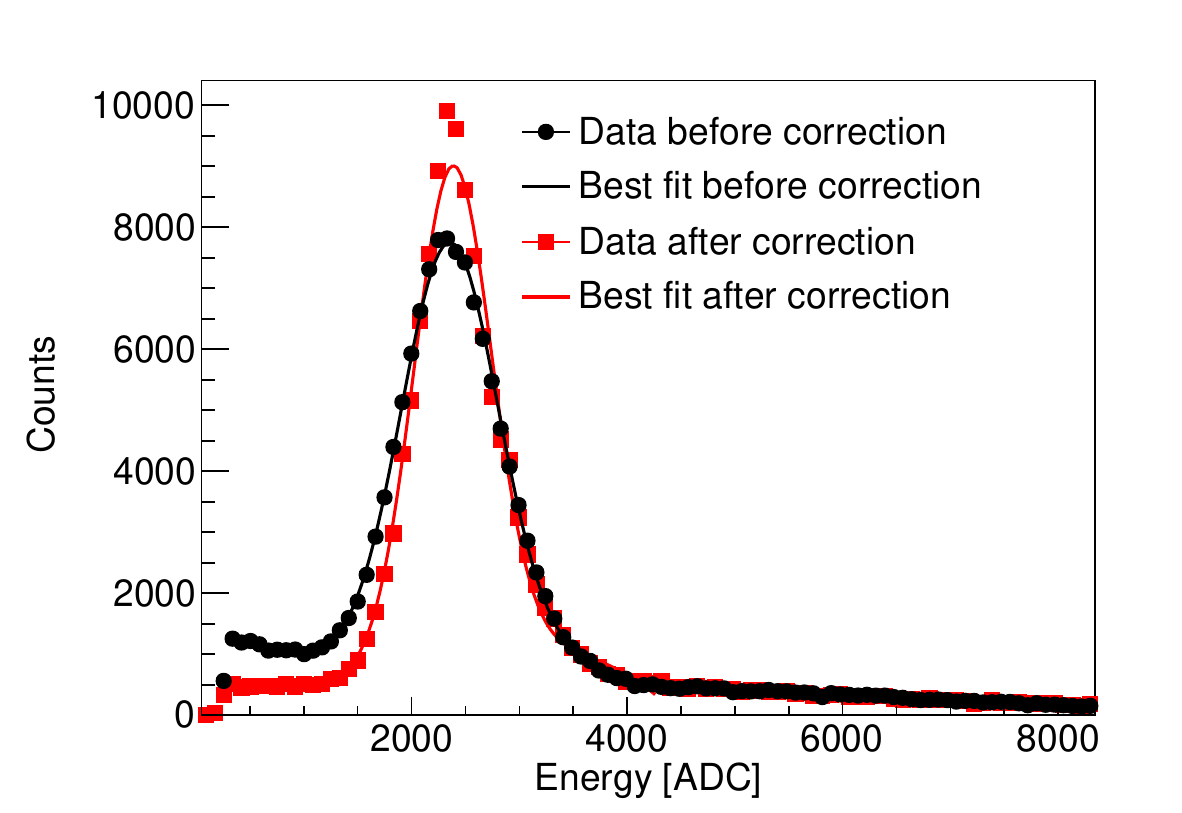}
    \caption{(Color online) An example of the energy spectrum before and after uniformity correction at 7 bar.}
    \label{spe_correction}
\end{figure}

\begin{figure}[!htbp]
    \includegraphics[width=\hsize]{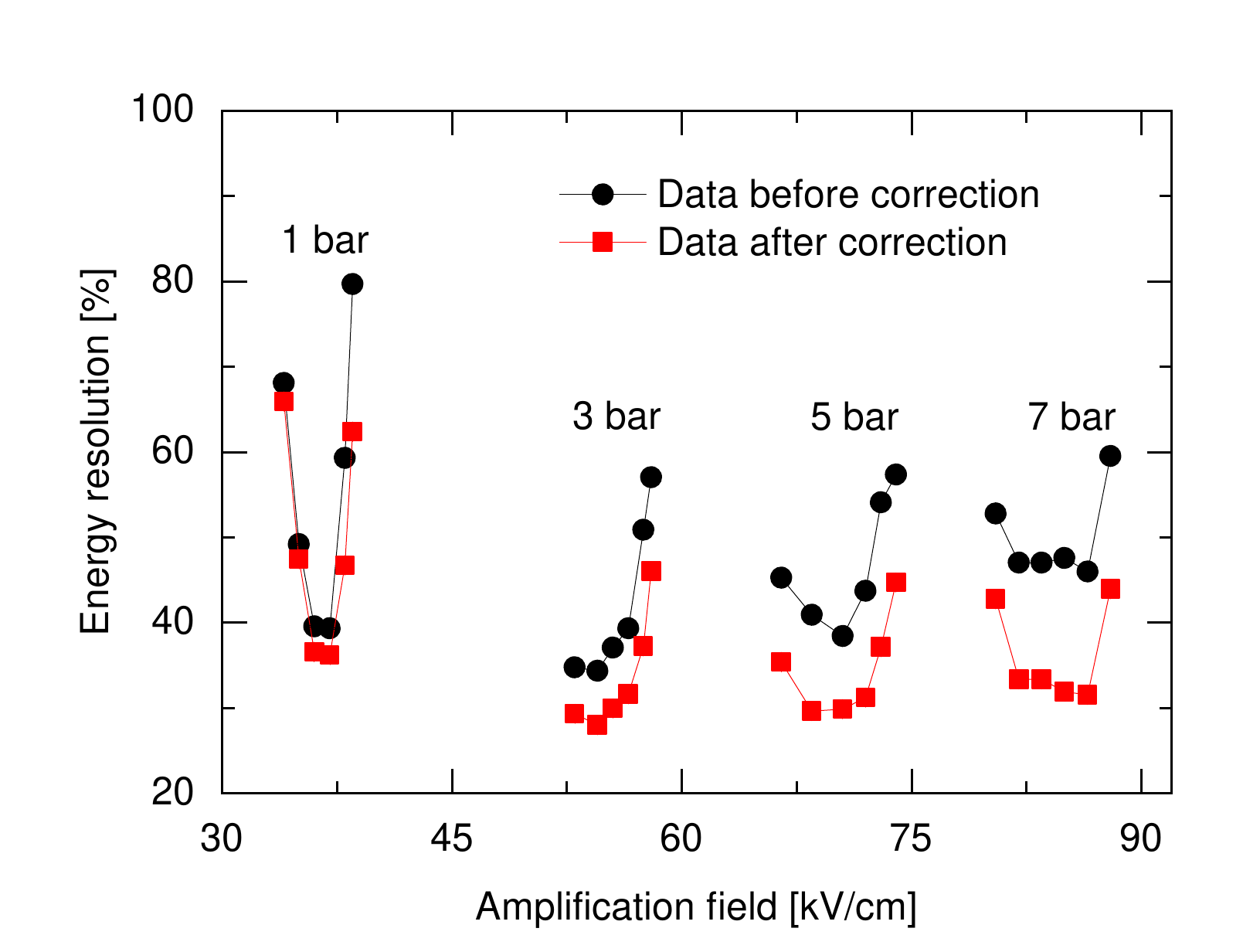}
    \caption{(Color online) The energy resolution before and after uniformity correction from 1 to 7 bar.}
    \label{resolutionAndCorrectResolution}
\end{figure}

The gain map can also be applied to the charge reconstruction of source events for uniformity correction under the same operating conditions.
Fig.~\ref{spe_correction} illustrates an example of the energy spectrum before and after uniformity correction at 7 bar.
After uniformity correction, the energy resolution is improved from 44.9\% to 35.4\%.
Because the relative gain of each region can not be kept constant under different amplification fields and gas mixtures, the gain map should be calibrated specifically for each pressure and amplification field.
The energy resolution before and after uniformity correction from 1 to 7 bar is shown in Fig.~\ref{resolutionAndCorrectResolution}.
The amplification field varies at each pressure in the measurements.
The electron clusters of \ce{^{37}Ar} decay events become smaller under higher pressures, which improves the efficiency of uniformity correction.
In general, the small-sized electron clusters deposited by particles lead to fewer triggered strips in the readout plane, which means that the detected charge is more representative of the gain at the hit position, and the gain maps used for uniformity correction are more efficient.

\subsection{Drift field evolution}

\begin{figure*}[htbp]
    \centering
    \subfigure[]{
        \label{subfig:hitmap1kv}
        \includegraphics[width=.36\hsize,trim=30 50 0 70, clip]{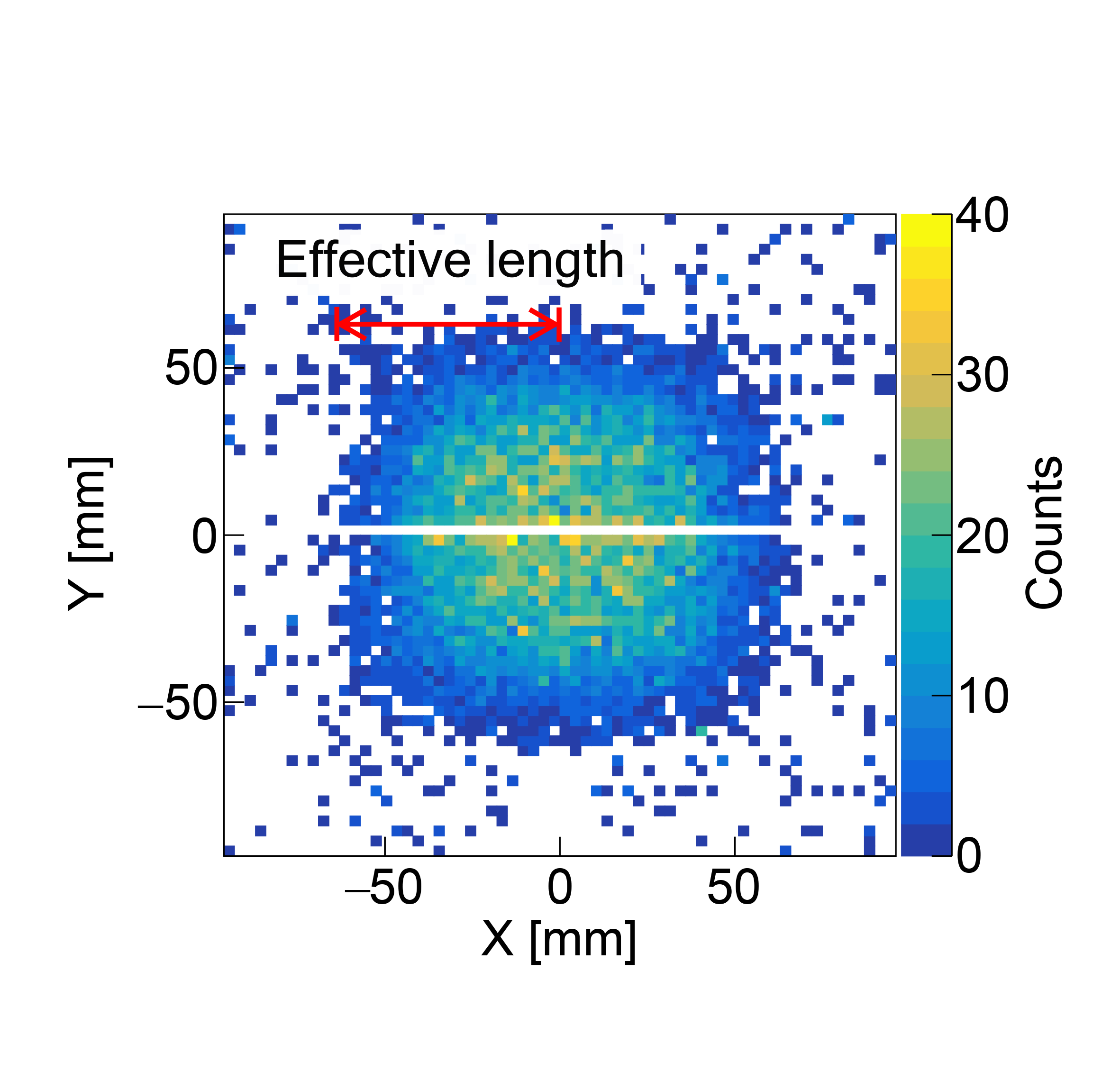}
 }
    \subfigure[]{
        \label{subfig:hitmap5kv}
        \includegraphics[width=.36\hsize,trim=30 50 0 70,clip]{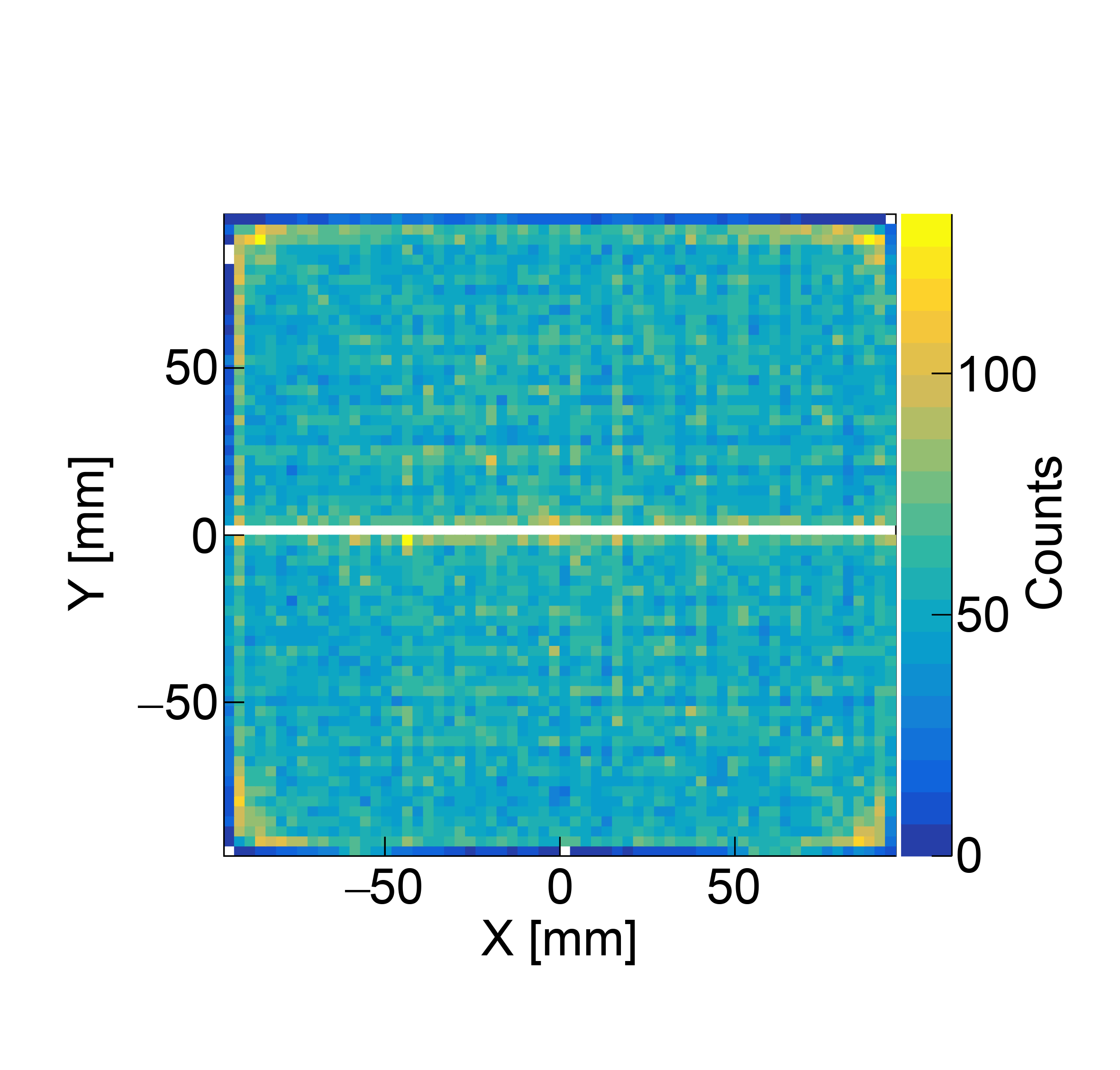}
 }
    \subfigure[]{
    \label{subfig:sim1kv}
        \includegraphics[width=.36\hsize,trim=0 0 142 0,clip]{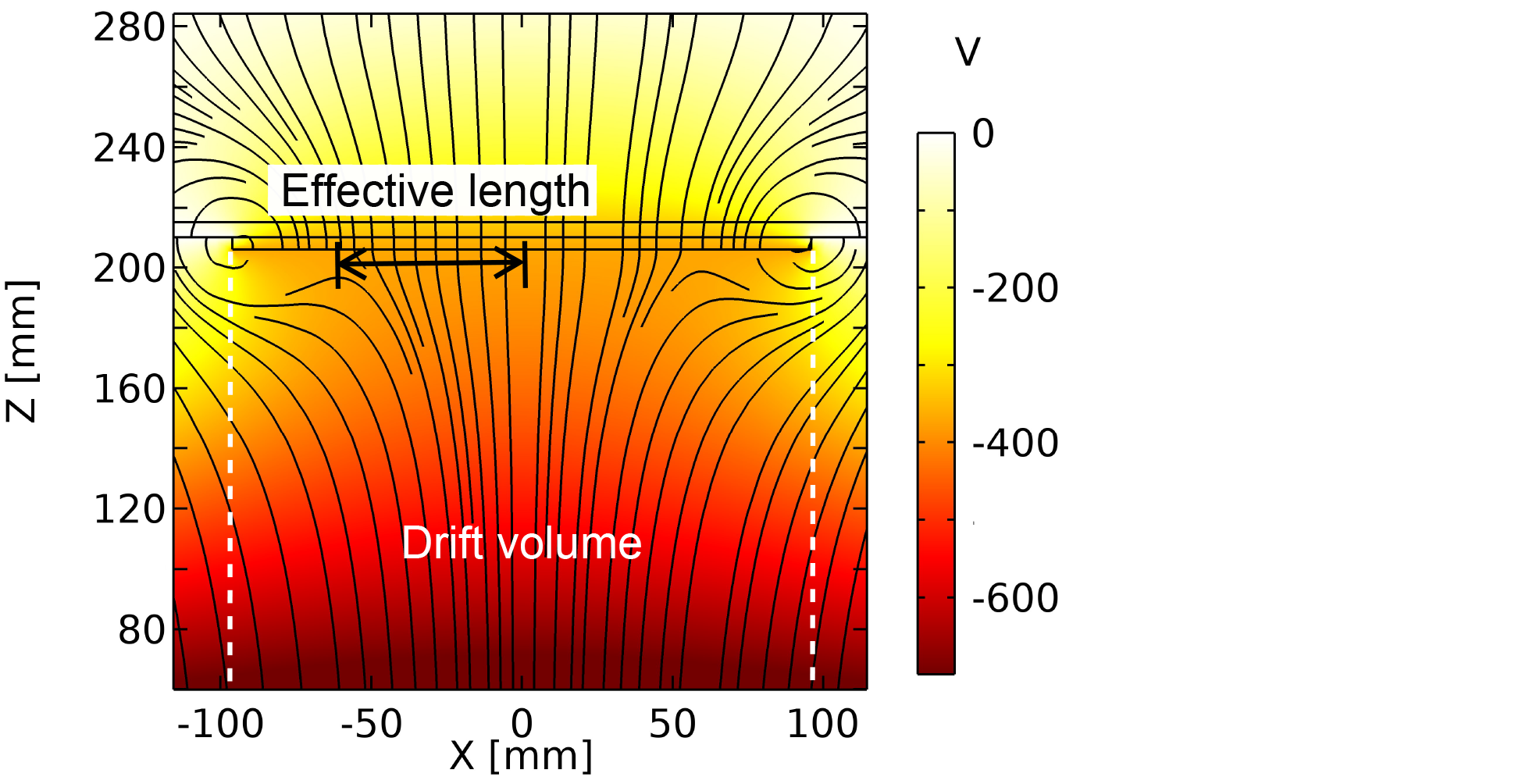}
 }
    \subfigure[]{
        \label{subfig:sim5kv}
        \includegraphics[width=.36\hsize,trim=0 0 142 0,clip]{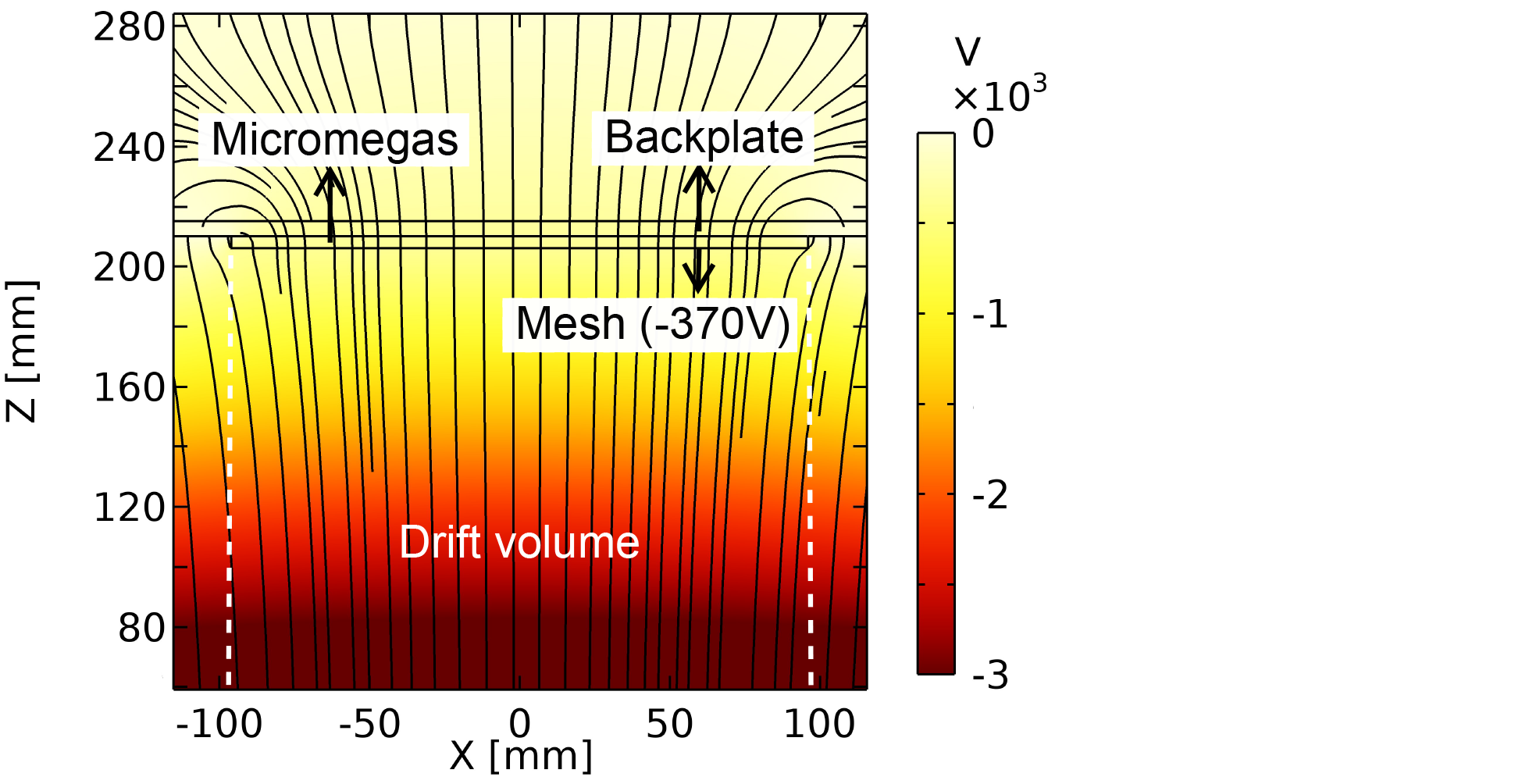}
 }
    \caption{\label{fig:DriftEvolution} 
 (a) A hit position map of source events in the XY readout plane at a drift field of 31.5~V/cm. 
 (b) A hit position map of source events in the XY readout plane at a drift field of 231.5~V/cm. 
 (c) Vertical electric field cross-section of the detector at a drift field of 31.5~V/cm simulated with COMSOL. 
 (d) Vertical electric field cross-section of the detector at a drift field of 231.5~V/cm simulated with COMSOL.}
\end{figure*}

A uniformly distributed $^{37}$Ar source provides an effective quantification of drift field distortion in the TPC.
The distortion causes electron clusters to deviate from their straight drift path or to run out of the readout plane entirely~\cite{bib:21}.
Fig.~\ref{subfig:hitmap1kv} and Fig.~\ref{subfig:hitmap5kv} show the hit position maps at 1 bar with drift voltages of -1~kV (a drift field of 31.5~V/cm) and -5~kV (a drift field of 231.5~V/cm), respectively.
The mesh voltage is fixed at -370~V, corresponding to an amplification field of 37~kV/cm.
To quantify the electric field distortion in the experiment, as shown in Fig.~\ref{subfig:hitmap1kv}, the effective charge collection length (referred to as effective length) is defined as the distance from the strip where the event rate is less than 1/10 of the highest value to the readout center.
The electric field distortion caused by the low drift field affects the collection efficiency, resulting in a dead zone at the edge of Micromegas.
Therefore, the effective length at a low drift field of 31.5~V/cm is significantly smaller than that at a high drift field of 231.5~V/cm.

The three-dimensional electrostatic field simulation of the detector is performed with the COMSOL Multiphysics software, which describes the realistic detector geometry, materials, and electric configurations.
A voltage of -1~kV and -5~kV is applied to the cathode to generate a drift field of 31.5~V/cm and 231.5~V/cm respectively, where the copper ring closest to the readout plane is set to 0~V.
The Micromegas mesh is set to a constant voltage of~-370 V, and the adjacent aluminum backplate on the upper side is grounded.
The vertical electric field cross-sections (i.e. the XZ plane) of the detector at electric fields of 31.5~V/cm and 231.5~V/cm are shown in Fig.~\ref{subfig:sim1kv} and Fig.~\ref{subfig:sim5kv}, respectively.
At a low drift field of 31.5~V/cm, the electric field distortion is pronounced at the edge of Micromegas, and the electric field lines bend at the edge.
As shown in Fig.~\ref{subfig:sim1kv}, the effective length in the simulation is defined as the distance from the inflection point where the electric field lines bend at the Micromegas edge to the readout center in the X(Y) direction.
Near the edges of the readout active area, the electric field line density is low, which corresponds to dead zones in the detector.
In the case of a drift field of 231.5~V/cm, the electric field distortion is negligible at the Micromegas edges, and thus the entire readout plane is effective for event collection.
The drift field distortion does not take into account charge accumulation in the TPC since the event rates in our detector are rather small.  

\begin{figure}[htb]
    \includegraphics[width=\hsize]{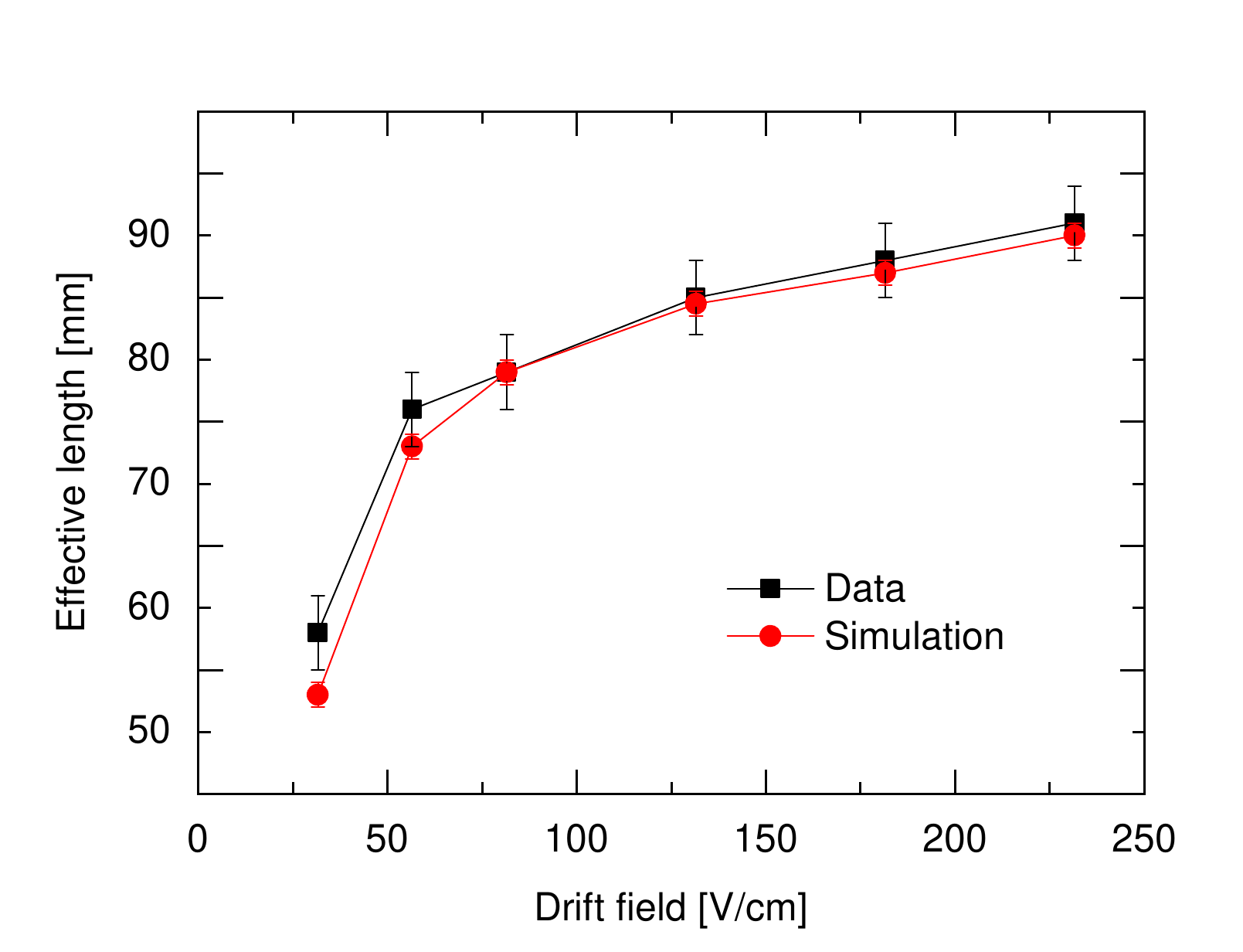}
    \caption{(Color online) Dependence of the effective length on the drift field.}
    \label{sensitiveLenVsDrift}
\end{figure}

As shown in Fig.~\ref{sensitiveLenVsDrift}, we plotted the evolution of effective length with the drift field using the experimental data and the simulation to quantify the distortion effect.
The experimental data have a 3~mm uncertainty (equal to the Micromegas strip pitch), while the COMSOL simulation has a 1~mm uncertainty due to solver precision.
Both experimental data and simulations show a non-linear positive correlation between the effective length and drift field strength.
Below an electric field of 56.5~V/cm, the effective length drops sharply with the decrease of the drift field.
At a low electric field of 31.5~V/cm, source events can be collected in the region about 6.3~cm away from the readout center, and the peripheral region becomes a dead zone due to the electric field distortion.
The effective length at a high electric field of 231.5~V/cm is about 9.6~cm, covering almost the entire sensitive area of the detector.

\section{Conclusions} \label{sec.V}

In summary, an internal calibration with \ce{^{37}Ar} is carried out to calibrate a gaseous TPC. 
The \ce{^{37}Ar} source is injected directly into the TPC and distributed uniformly within the active volume. 
The fast-decay, low-energy-deposition, and uniform-distribution features of \ce{^{37}Ar} enable us to efficiently investigate the detector performance, including electron transmission, gain, energy resolution, gain uniformity, and drift field evolution.
The electron transmission is first scanned at pressures from 1 to 10 bar, and the plateau is selected to ensure the most optimized electron transparency for this detector configuration.
The maximum attainable gain reaches up to more than ten thousand in most cases. 
The limiting factor for a higher gain value is the micro-discharges between the mesh and the anode, especially at a low pressure of 0.3 bar and a high pressure of 10 bar.
An optimal energy resolution of 33.4\% FWHM is achieved at 2.82 keV in a 3~bar detector gas, with a gain of $5.1\times 10^3$.
The gain uniformity is also calibrated with \ce{^{37}Ar} decay events, and a non-uniformity of 13.0\% (RMS/Mean) is obtained with an amplification field of 37~kV/cm at 1 bar.
The obtained gain map is used to assess the inhomogeneity of the avalanche gap of Micromegas modules.
The detected charge amplitude at each readout strip is corrected using the gain uniformity map to obtain an improved energy resolution.
Since the electron clusters of \ce{^{37}Ar} decay events are more concentrated under high pressures, the uniformity correction is more impactful, and the energy resolution is considerably improved with this method.
Particularly at 7 bar, the energy resolution is enhanced from 44.9\% to 35.4\% after the uniformity correction is applied.
Because of the severe electric field distortion caused by the low drift field, the effective length under the low drift field is significantly smaller than that under the high drift field. 
Our tests also observe the effective length of the detector evolving with the drift field, which is verified by the electric field simulation using COMSOL.

The \ce{^{37}Ar} internal calibration method offers distinct advantages over the traditionally used \ce{^{55}Fe} source.
A gaseous source can be easily injected or removed from the detector, which greatly simplifies the calibration procedure and reduces the turn-around time for physics experiments involving gaseous TPCs.
The internal source also makes \emph{in situ} calibration at low and high pressures possible. 
Detector performance calibrated at the same operating conditions as physics runs reduces systematic uncertainties in data analysis.
For detectors that are equipped with large-area readout modules such as Micromegas, \ce{^{37}Ar} calibration emerges as an excellent and preferred choice for efficiently calibrating the entire detector readout plane.
In addition, we are currently developing particle trajectory recognition algorithms with the \ce{^{37}Ar} source, which can be used for the calibration of dark matter directional detection detectors.
There are a couple of minor concerns for \ce{^{37}Ar}, both of which can be easily mitigated.  
Due to the short half-life, the shell life of \ce{^{37}Ar} sources is limited and regular source production is desired. 
Safety issues related to a gaseous source, such as leak detection and radiation monitoring, should be addressed in the experiment.

\section*{Acknowledgment}

The authors declare that they have no known conflicts of interest in terms of competing financial interests or personal relationships that could have an influence or are relevant to the work reported in this article.

\end{document}